\documentclass[aps,prd,twocolumn,superscriptaddress,nofootinbib,preprintnumbers]{revtex4-1}
\usepackage{amsmath}
\usepackage{graphicx}
\usepackage{subfigure}
\usepackage{amssymb}
\usepackage{xcolor}
\usepackage{multirow}
\usepackage{cancel}
\usepackage{color}
\usepackage{ulem}
\usepackage{listings}
\usepackage{xcolor}
\usepackage{float}
\usepackage{slashed}

\usepackage{tikz}
\usetikzlibrary{arrows,shapes}
\usetikzlibrary{trees}
\usetikzlibrary{matrix,arrows} 				
\usetikzlibrary{positioning}				
\usetikzlibrary{calc,through}				
\usetikzlibrary{decorations.pathreplacing}  
\usepackage{pgffor}							

\usetikzlibrary{decorations.pathmorphing}	
\usetikzlibrary{decorations.markings}
\tikzset{
    vector/.style={decorate, decoration={snake}, draw},
	provector/.style={decorate, decoration={snake,amplitude=2.5pt}, draw},
	antivector/.style={decorate, decoration={snake,amplitude=-2.5pt}, draw},
    fermion/.style={draw=black, postaction={decorate},
        decoration={markings,mark=at position .55 with {\arrow[draw=black]{>}}}},
    fermionbar/.style={draw=black, postaction={decorate},
        decoration={markings,mark=at position .55 with {\arrow[draw=black]{<}}}},
    fermionnoarrow/.style={draw=black},
    gluon/.style={decorate, draw=black,
        decoration={coil,amplitude=4pt, segment length=5pt}},
    scalar/.style={dashed,draw=black, postaction={decorate},
        decoration={markings,mark=at position .55 with {\arrow[draw=black]{>}}}},
    scalarbar/.style={dashed,draw=black, postaction={decorate},
        decoration={markings,mark=at position .55 with {\arrow[draw=black]{<}}}},
    scalarnoarrow/.style={dashed,draw=black},
    electron/.style={draw=black, postaction={decorate},
        decoration={markings,mark=at position .55 with {\arrow[draw=black]{>}}}},
	bigvector/.style={decorate, decoration={snake,amplitude=4pt}, draw},
}

\tikzstyle{block} = [draw, rectangle, 
    minimum height=3em, minimum width=6em]

\usepackage[colorlinks,citecolor=blue]{hyperref}
\usepackage{amsmath}
\usepackage{wrapfig}

\newcommand{\be}{\begin{equation}}
\newcommand{\ee}{\end{equation}}
\newcommand{\beq}{\begin{equation}}
\newcommand{\eeq}{\end{equation}}
\newcommand{\bea}{\begin{eqnarray}}
\newcommand{\eea}{\end{eqnarray}}
\newcommand{\besp}{\begin{equation}\begin{split}}
\newcommand{\eesp}{\end{split}\end{equation}}

\newcommand{\nn}{\nonumber}

\newcommand{\Eq}[1]{Eq.~(\ref{#1})}

\newcommand{\Dfbd}{\mathord{\buildrel{\lower3pt\hbox{$\scriptscriptstyle\leftrightarrow$}}\over {D}_{\mu}}}
\newcommand{\ave}[1]{\left\langle #1\right\rangle}

\def\mL{\mathcal{L}}

\def\mO{\mathcal{O}}

\def\0{\textbf{0}}
\def\1{\textbf{1}}
\def\2{\textbf{2}}
\def\3{\textbf{3}}
\def\4{\textbf{4}}
\def\5{\textbf{5}}
\def\6{\textbf{6}}
\def\7{\textbf{7}}
\def\8{\textbf{8}}
\def\9{\textbf{9}}

\begin{document}

\title{Primordial black holes from a cosmic phase transition: The collapse of Fermi-balls}

\author{Kiyoharu Kawana}
\email{kawana@snu.ac.kr}
\affiliation{Center for Theoretical Physics, Department of Physics and Astronomy, Seoul National University, Seoul 08826, Korea}

\author{Ke-Pan Xie}
\email{Corresponding author. kepan.xie@unl.edu}
\affiliation{Department of Physics and Astronomy, University of Nebraska, Lincoln, NE 68588, USA}
\affiliation{Center for Theoretical Physics, Department of Physics and Astronomy, Seoul National University, Seoul 08826, Korea}

\begin{abstract}

We propose a novel primordial black hole (PBH) formation mechanism based on a first-order phase transition (FOPT). 
If a fermion species gains a huge mass in the true vacuum, the corresponding particles get trapped in the false vacuum as they do not have sufficient energy to penetrate the bubble wall. After the FOPT, the fermions are compressed into the false vacuum remnants to form non-topological solitons called Fermi-balls, and then collapse to PBHs due to the 
Yukawa attractive force. We derive the PBH mass and abundance, showing that for a $\mO({\rm GeV})$ FOPT the PBHs could be $\sim10^{17}$ g and explain all of dark matter. 
If the FOPT happens at higher scale, PBHs are typically overproduced and extra dilution mechanism is necessary to satisfy current constraints. 

\end{abstract}

\maketitle

\section{Introduction}

Primordial black holes (PBHs) are hypothetical black holes which form prior to any galaxies and stars~\cite{zel1967hypothesis,hawking1971gravitationally}. 
Although not yet confirmed by experiments, PBHs have been a source of interest, as they can serve as a good candidate for dark matter (DM)~\cite{hawking1971gravitationally,Chapline:1975ojl,Khlopov:2008qy,Carr:2016drx,Carr:2020gox,Carr:2020xqk,Green:2020jor}, can seed supermassive black holes~\cite{Bean:2002kx,Khlopov:2004sc,Duechting:2004dk,Kawasaki:2012kn,Clesse:2015wea}, can generate the baryon asymmetry of the Universe~\cite{Hawking:1974rv,carr1976some,Baumann:2007yr,Hook:2014mla,Fujita:2014hha,Hamada:2016jnq,Morrison:2018xla,Hooper:2020otu,Perez-Gonzalez:2020vnz,Datta:2020bht,JyotiDas:2021shi}, and can explain some gravitational wave (GW) signals at LIGO/Virgo~\cite{Abbott:2016blz,Abbott:2016nmj,Abbott:2017vtc,Clesse:2016vqa,Bird:2016dcv,Sasaki:2016jop}, etc. 
 While the most popular PBH formation mechanism is the collapse of the overdense region from primordial perturbations of inflation~\cite{Carr:1974nx,carr1975primordial,Sasaki:2018dmp}, there are other scenarios such as the collapse of cosmic topological defects~\cite{Hawking:1987bn,Caldwell:1995fu,Garriga:1992nm,Rubin:2000dq,Rubin:2001yw,Dokuchaev:2004kr,Deng:2016vzb}, scalar field fragmentation~\cite{Cotner:2016cvr,Cotner:2017tir,Cotner:2018vug,Cotner:2019ykd}, etc. PBHs can also form during a first-order phase transition (FOPT) in the early Universe via bubble collisions~\cite{Crawford:1982yz,Hawking:1982ga,La:1989st,Moss:1994iq,konoplich1998formation,Konoplich:1999qq,Kodama:1982sf,Kusenko:2020pcg,Baker:2021nyl}.

In this letter, we propose a novel PBH formation mechanism based on the collapse of non-topological solitons produced during a cosmic FOPT.
The simplest realization of this mechanism consists of a real scalar $\phi$ and a Dirac fermion $\chi$, with the Lagrangian
\be\label{L}
\mL=-\frac12\partial_\mu\phi\partial^\mu\phi-U(\phi)+\bar\chi i\slashed{\partial}\chi-g_\chi\phi\bar\chi\chi~, 
\ee
which conserves the fermion number by a global $U(1)_Q$. The scalar potential $U(\phi)$ triggers a FOPT from $\ave{\phi}=0$ to $w_*$ at temperature $T_*$.
If $g_\chi w_*\gg T_*$, the fermions cannot penetrate into the new (true) vacuum bubbles, where they acquire a mass $M_\chi^*\equiv g_\chi w_*$ that significantly exceeds their thermal kinetic energy. 
Consequently, after the FOPT, fermions are trapped in the old (false) vacuum and compressed to form non-topological solitons called {\it Fermi-balls}~\cite{Hong:2020est} if there is an asymmetry between the number densities of $\chi$ and $\bar\chi$, such that only $\chi$'s survive the $\chi\bar\chi\to\phi\phi$ annihilation. 
The conditions for Fermi-ball formation can be easily satisfied in many new physics models~\cite{Hong:2020est}: $g_\chi w_*\gg T_*$ can be realized by a supercooled FOPT~\cite{Creminelli:2001th,Nardini:2007me,Konstandin:2011dr,Jinno:2016knw,Marzo:2018nov,Hambye:2018qjv,Baratella:2018pxi} or strong coupling~\cite{Carena:2004ha,Angelescu:2018dkk} while the $\chi$ asymmetry can be generated by various asymmetric DM mechanisms~\cite{Kaplan:2009ag,Petraki:2013wwa,Zurek:2013wia}.

Fermi-balls are macroscopic compact objects that collect huge $Q$-charge originating from the fermion asymmetry. 
Inside the Fermi-ball is the false vacuum $\ave\phi=0$, in which $\chi$'s are massless fermions interacting with each other via the attractive Yukawa potential
\be
V(r)=-\frac{g_\chi^2}{4\pi r}e^{-M_\phi r},
\label{yukawa potential}
\ee
whose range of force $M_\phi^{-1}$ is determined by the effective mass at the false vacuum
\be\label{phi mass}
M_\phi^2=\frac{d^2U(\phi,T)}{d\phi^2}\Big|_{\phi=0}=\mu^2+c\,T^2,
\ee
where $U(\phi,T)$ is the thermal potential, $c$ is the thermal coefficient contributed by the light degrees of freedom (DOF) in a model. 
At the beginning, the effect of Eq.~(\ref{yukawa potential}) is negligible because its range of force $M_\phi^{-1}$ is very small, and we can treat $\chi$'s as independent particles obeying the Fermi-Dirac distribution.
However, $M_\phi^{-1}$ increases as the Universe cools down. When $M_\phi^{-1}$ is comparable to the mean separation of $\chi$'s inside the Fermi-ball, the attractive Yukawa force dominates and Fermi-balls collapse into PBHs.

\begin{figure*}
\centering
\subfigure{
\includegraphics[scale=0.25]{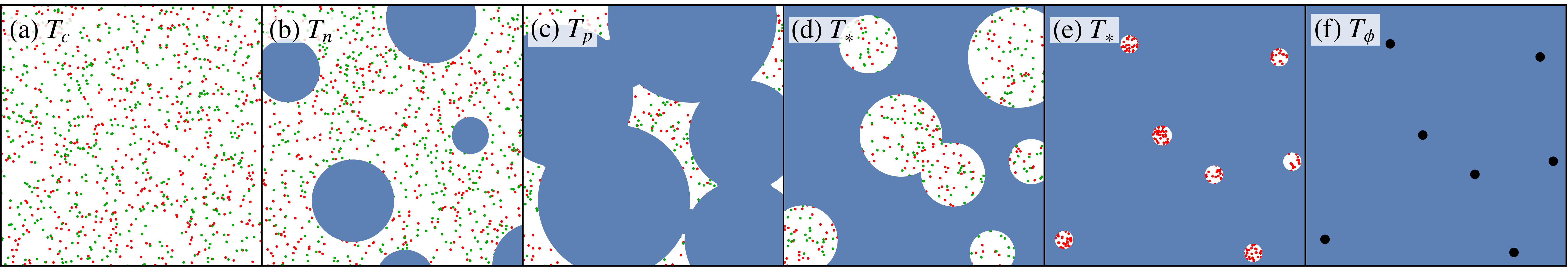}}
\caption{Sketch of the mechanism. {\bf (a)} $T_c$: $\chi$ (red points) and $\bar\chi$ (green points) live in the false vacuum (white). {\bf (b)} $T_n$: true vacuum bubbles (blue) nucleate. Fermions cannot penetrate into the bubbles due to the large mass gap. {\bf (c)} $T_p$: Fermions are trapped in the false vacuum remnants. {\bf (d,e)} $T_*$: remnants shrink to form Fermi-balls. {\bf (f)} $T_\phi$: Fermi-balls collapse into PBHs.}
\label{fig:sketch}
\end{figure*}

In short, the FOPT forms Fermi-balls, which in turn collapse into PBHs when the internal Yukawa force becomes dominant.\footnote{The collapse of Fermi-balls via gravity is recently studied in~\cite{Gross:2021qgx}.
} 
PBH formation via the attractive scalar force has been also studied in Refs.~\cite{Amendola:2017xhl,Flores:2020drq} where  an ultralight scalar is introduced to provide a long-range force, causing the growth of density perturbation in the plasma. 
In our work, $\phi$ needs not to be light: the fermions inside a Fermi-ball are so dense that even a relatively short-range Yukawa force is able to cause instability, and collapse happens only in individual Fermi-balls.
Since this situation can be easily realized in many particle physics models, we conclude that our scenario is rather generic. 
The mechanism is illustrated schematically in Fig.~\ref{fig:sketch}, and the details are discussed below. 

\section{Fermi-ball formation in FOPT}

At critical temperature $T_c$, the two vacua are degenerate, and the Universe still stays in the false vacuum $\ave\phi=0$, see Fig.~\ref{fig:sketch}a. 
Below $T_c^{}$, the Universe acquires a probability of decaying to  the true vacuum $\ave\phi=w(T)$ that has a lower energy, and the decay rate is dominated by the $O(3)$-symmetric bounce action $S_3(T)$ of the thermal potential $U(\phi,T)$~\cite{Linde:1981zj}; 
\be
\Gamma(T)\sim T^4e^{-S_3(T)/T}~.
\ee
Bubbles start to nucleate at $T_n^{}$ when
\be
\int_{T_n}^{T_c}\frac{dT}{T}\frac{\Gamma(T)}{H^{4}(T)}\approx1,
\ee
(see Fig.~\ref{fig:sketch}b), where the Hubble constant is $H(T)^2=(8\pi/3M_{\rm Pl}^2)\left(\rho_R(T)+\rho_U(T)\right)$. 
Here $\rho_R(T)=\pi^2g_*T^4/30$ is the radiation energy density,  $g_*^{}$ is the number of relativistic DOF,  $\rho_U(T)=U(0,T)-U(w,0)$ is the energy of the false vacuum with respect to the true vacuum at $T=0$, and $M_{\rm Pl}^{}=1.22\times10^{19}$~GeV.

Given $\Gamma(T)$ and $H(T)$, we can define the the volume fraction of the false vacuum to the Universe as $p(T)=e^{-I(T)}$ where $I(T)$ is the amount of true vacuum volume per unit comoving volume~\cite{Guth:1981uk,Ellis:2019oqb,Wang:2020jrd}. 
The percolation temperature $T_p$ is defined by $p(T_p)=0.71$~\cite{rintoul1997precise} at which the bubbles form an infinite connected cluster, see Fig.~\ref{fig:sketch}c. 
The Fermi-balls form at $T_*^{}$ defined by $p(T_*)=
0.29$~\cite{Hong:2020est} at which the false vacuum is separated into disconnected remnants, which first split and then shrink to Fermi-balls. 
The critical remnant is defined as a remnant that just ends splitting and starts to shrink, and its radius $R_*$ is determined by \cite{Hong:2020est}
\be\label{right_after}
\Gamma(T_*)V_*\left(\frac{R_*}{v_b}\right)\sim1~,\quad V_*=\frac{4\pi}{3}R_*^3~,
\ee
where $v_b$ is the bubble wall velocity. 
\Eq{right_after} means that the critical remnant shrinks to a Fermi-ball before another true vacuum bubble is created inside it. The number density of the critical remnants is given by $n_{\rm rem.}^*=V_*^{-1}p(T_*)$ and it is also the number density of Fermi-balls right after the formation, $n_{\rm FB}^*=n_{\rm rem.}^*$.

Fermions are trapped in the false vacuum and the trapping fraction $F_\chi^{\rm trap.}$ is a function of $v_b$ and $M_\chi^*/T_*=g_\chi w_*/T_*$~\cite{Hong:2020est,Chway:2019kft}, where $w_*=w(T_*)$. For example, $M_\chi^*=12\,T_*$ and $v_b=0.6$ yield $F_\chi^{\rm trap.}\approx0.98$ in a benchmark model \cite{Hong:2020est}.\footnote{More precisely, an FOPT by the generic singlet scalar potential $\mu^2\phi^2/2+\mu_3\phi^3/3+\lambda\phi^4/4$ was studied in Ref.~\cite{Hong:2020est}. In such a model, it is possible to realize $w_*/T_*\sim\mO(10)$ if $|\mu_3|$ is fairly large.
} 
When the remnants shrink, the fermions are forced to annihilate via $\chi\bar\chi\to\phi\phi$. The $\phi$ bosons are free to pass through the wall to the true vacuum, and finally annihilate/decay to the Standard Model (SM) particles via the Higgs portal coupling $\phi^2|H|^2$. Describing the $\chi$ asymmetry by $\eta_\chi\equiv (n_\chi-n_{\bar\chi})/s(T)$ with $s(T)=2\pi^2g_*T^3/45$ being the entropy density, the number of $\chi$ fermions surviving the annihilation in a critical remnant is
\be
Q_{\rm FB}=F_\chi^{\rm trap.}\frac{n_\chi-n_{\bar\chi}}{n_{\rm FB}^*}=F_\chi^{\rm trap.}\frac{\eta_\chi s_*V_*}{p(T_*)}~,
\ee
where $s_*=s(T_*)$. 
A remnant stops shrinking when the trapped fermions' degeneracy pressure is able to balance the vacuum pressure $U_0(T_*)$. 
Once such balance is built, the remnant together with its trapped fermions form a Fermi-ball, see Fig.~\ref{fig:sketch}d and e.

Let us now evaluate the profile of a Fermi-ball.
Minimizing the Fermi-ball energy
\be\label{FB_E}
E_{\rm FB}=\frac{3\pi}{4}\left(\frac{3}{2\pi}\right)^{2/3}\frac{Q_{\rm FB}^{4/3}}{R}+\frac{4\pi}{3}U_0(T_*)R^3,
\ee
yields the mass and radius as~\cite{Hong:2020est}\footnote{The FOPT-induced Fermi-ball is first proposed in Ref.~\cite{Hong:2020est}; the terminology ``Fermi-ball'' is also used in literatures with different physical meanings~\cite{Macpherson:1994wf,Macpherson:1994vk,Sivaram:2011jk}.
There are various DM mechanisms based on trapping particles into the false vacuum: quark nuggets~\cite{Witten:1984rs,Frieman:1990nh,Zhitnitsky:2002qa,Oaknin:2003uv,Lawson:2012zu,Atreya:2014sca,Bai:2018vik,Bai:2018dxf}, accidentally asymmetric DM~\cite{Asadi:2021yml}, FOPT-induced Q-balls~\cite{Krylov:2013qe,Huang:2017kzu}, etc.}
\be\label{MR_profile}
M_{\rm FB}=Q_{\rm FB}\left(12\pi^2U_0(T_*)\right)^{1/4},~
R_{\rm FB}^3=\frac{3}{16\pi}\frac{M_{\rm FB}}{U_0(T_*)}~,
\ee
where $U_0(T)\equiv U(0,T)-U(w(T),T)$ is the vacuum energy difference between the interior and exterior of the Fermi-ball. Note that the first term in \Eq{FB_E} only considers the degeneracy pressure of the Fermi gas and we neglect the subdominant contribution from thermal excitations.  
See the appendix for a more accurate Fermi-ball profile.
As we will see, $R_{\rm FB}$ is typically much larger than the Schwarzschild radius of a Fermi-ball, thus gravity can be neglected in the calculation. 
Recently, it is proposed that during a FOPT the trapped fermions might be compressed to be a PBH without forming any stable soliton in Ref.~\cite{Baker:2021nyl}. 
Numerical simulations show that such a scenario requires a very large mass gap $M_\chi^*/T_*\gtrsim200$ and small Yukawa $g_\chi\lesssim10^{-3}$~\cite{Baker:2021nyl}. 
Since we consider $M_\chi^*/T_*\sim\mO(10)$ and $g_\chi\sim\mO(1)$ in this letter, Fermi-balls can safely form during the FOPT in our scenario.

\section{Collapse of Fermi-balls into PBHs}

\begin{figure}
\centering
\subfigure{
\includegraphics[scale=0.32]{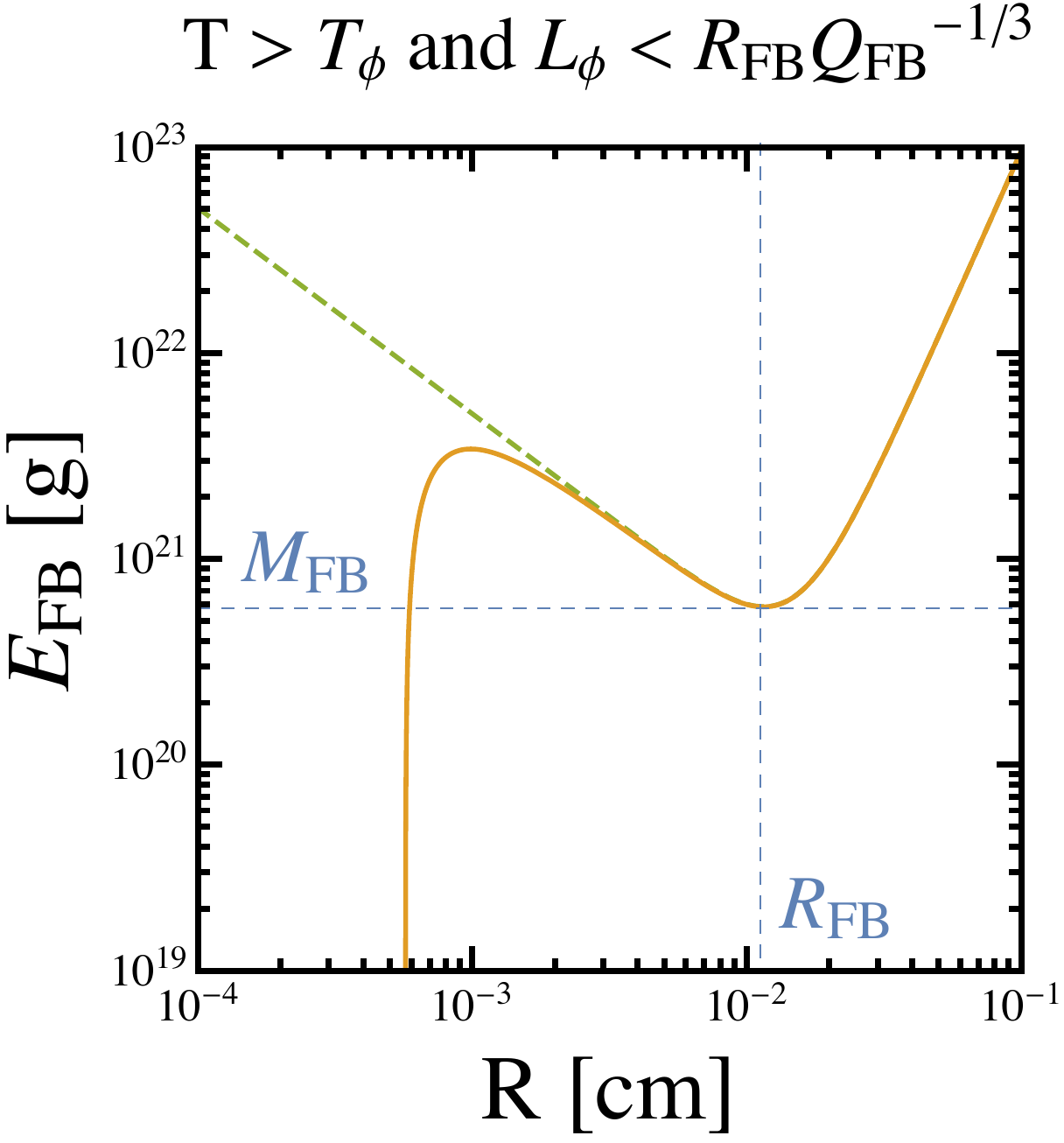}}\quad
\subfigure{
\includegraphics[scale=0.32]{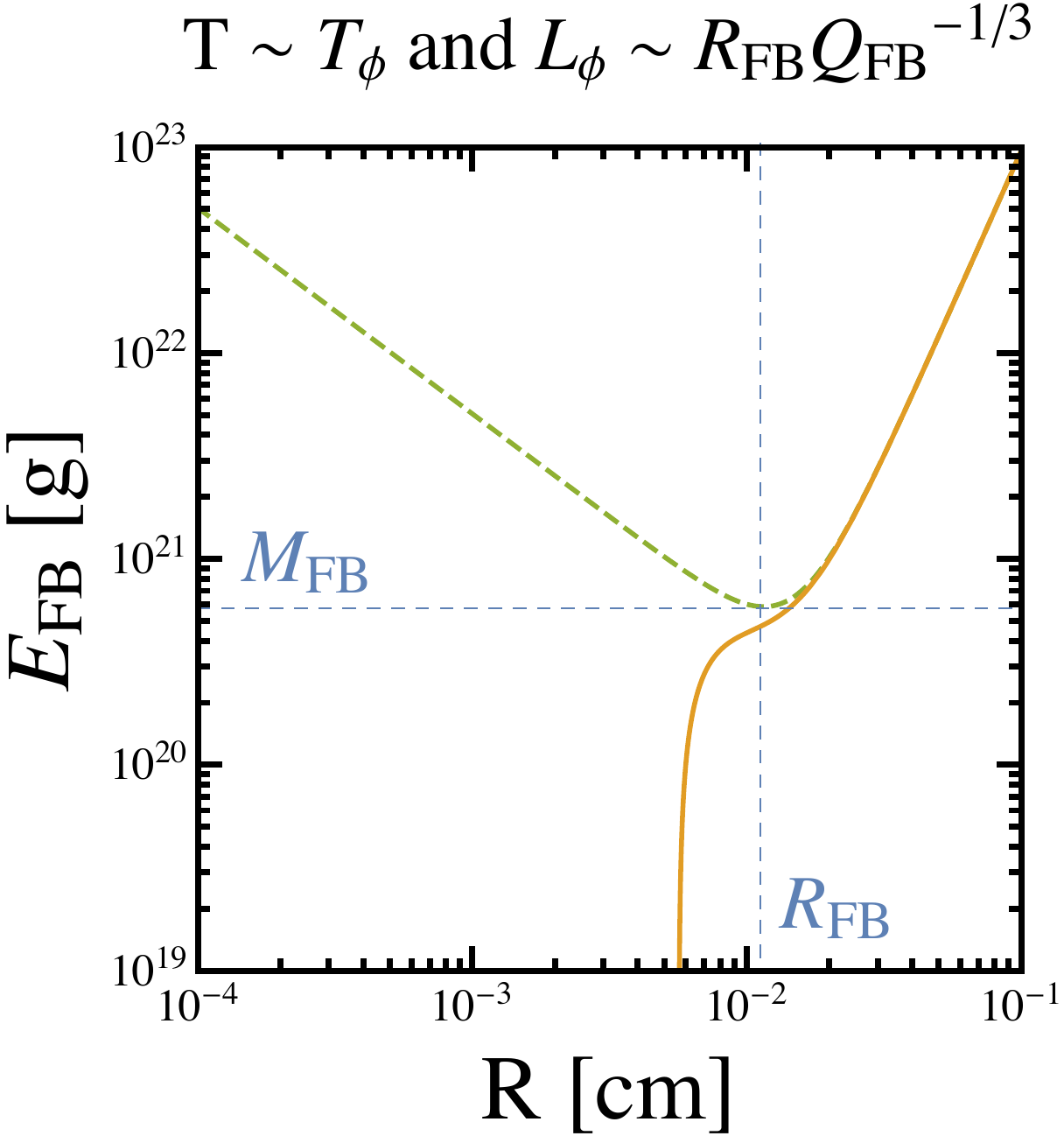}}
\caption{The energy profile (orange lines) changes as $T$ drops. Left: when $T>T_\phi$, the range of Yukawa force $L_\phi$ is small compared with the mean separation $R_{\rm FB}Q_{\rm FB}^{-1/3}$, and hence a stable Fermi-ball solution can exist. The green dashed line represents the energy profile without the Yukawa interaction. Right: when $T$ cools down to $T_\phi$, $L_\phi\sim R_{\rm FB}Q_{\rm FB}^{-1/3}$ and causes the instability. In that case there is no stable solution any more and the Fermi-ball collapses into a PBH.}
\label{fig:profile}
\end{figure}

After formation, the Fermi-ball cools down by emitting light particles, e.g. $\chi\to\chi f\bar f$ where $f$ could be electrons or neutrinos~\cite{Witten:1984rs} (via the $\phi^2|H|^2$ or other portal couplings). As shown in the appendix, the cooling time scale of the Fermi-ball is much shorter than the Universe expansion. 
Therefore, the Fermi-ball is able to track the temperature of the plasma, so that its profile changes slowly by replacing $T_*\to T$ in \Eq{MR_profile}, and the number density is diluted as $n_{\rm FB}=n_{\rm FB}^*s(T)/s_*$. 
As the temperature drops, the Yukawa potential Eq.~(\ref{yukawa potential}) starts to play a role because its range of force inside the Fermi-ball $L_\phi\equiv M_\phi^{-1}=(\mu^2+c\,T^2)^{-1/2}$ correspondingly increases. 
Then, we should add the Yukawa potential energy of a Fermi-ball into \Eq{FB_E}:
\be
\Delta E_{\rm FB}\approx-\frac{3g_\chi^2}{20\pi}\frac{Q_{\rm FB}^2}{R}\times\frac{5}{2}\left(\frac{L_\phi}{R}\right)^2~.
\ee
When $L_\phi/R_{\rm FB}\sim Q_{\rm FB}^{-1/3}$ (which means that  the correlation length reaches the mean separation of $\chi$'s), the above energy dominates over the Fermi gas kinetic term in Eq.~(\ref{FB_E}), and its negative sign causes the instability of Fermi-balls, resulting in the collapse into PBHs. See Fig.~\ref{fig:profile} and Fig.~\ref{fig:sketch}f for illustrations of the energy profile and PBH formation, respectively.
A more accurate expression of the Yukawa energy is given in the appendix.
The PBH mass and number density right after formation are given by $M_{\rm PBH}=M_{\rm FB}|_{T_*\to T_\phi}$ and $n_{\rm FB}^*s(T_\phi)/s_*$, respectively, where $T_\phi$ is the collapse temperature.

Up to now, our PBH scenario has been completely described. Given the thermal potential $U(\phi,T)$, one can derive $T_*$ by calculating $\Gamma(T)$ and $H(T)$. The Fermi-ball profile is determined by $\Gamma(T_*)$ and $U_0(T_*)$, while the PBH profile is derived by running the Fermi-ball profile to $T_\phi$. 
Below we give some general estimates for the PBH profile based on a few simplified assumptions.

\section{The PBH profile}

We consider the radiation dominated era and approximate $T_*\approx T_p$ since the Fermi-ball formation is very close to the percolation. 
In this case, the bounce action can be obtained as~\cite{Huber:2007vva}
\begin{multline}
\frac{S_3(T_*)}{T_*}\approx131-4\ln\left(\frac{T_*}{100~{\rm GeV}}\right)\\
-4\ln\left(\frac{\beta/H}{100}\right)+3\ln v_b-2\ln\left(\frac{g_*}{100}\right)~,
\end{multline}
where $\beta/H$ is the time scale ratio of the Universe expansion and the FOPT. 
Next, we assume $U_0(T_*)\approx\alpha\times\rho_R(T_*)$, where $\alpha$ is the ratio of the latent heat of FOPT to the radiation energy density.
The parameters $\alpha$ and $\beta/H$ are also crucial in the calculation of GWs~\cite{Grojean:2006bp,Caprini:2015zlo,Caprini:2019egz,Guo:2021qcq} from a FOPT.  
For simplicity, we 
omit the temperature dependence of the Ferm-ball mass. 

We are now able to express the Fermi-ball and PBH profiles as functions of $(v_b,\eta_\chi,T_*,\alpha,\beta/H)$. 
Let $F_\chi^{\rm trap.}\approx1$, the charge and radius of a Fermi-ball are given by
\begin{multline}
Q_{\rm FB}\approx1.0\times10^{42}\times v_b^3\left(\frac{\eta_\chi}{10^{-3}}\right)\times\\
\left(\frac{100}{g_*}\right)^{1/2}\left(\frac{100~{\rm GeV}}{T_*}\right)^3\left(\frac{100}{\beta/H}\right)^3~,
\end{multline}
\begin{multline}
R_{\rm FB}\approx 4.8\times10^{-3}~{\rm cm}
\times v_b\left(\frac{\eta_\chi}{10^{-3}}\right)^{1/3}\times\\
\left(\frac{100}{g_*}\right)^{5/12}\left(\frac{100~{\rm GeV}}{T_*}\right)^2\left(\frac{100}{\beta/H}\right)\alpha^{-1/4}~.
\end{multline}
Note that by definition $\eta_\chi\lesssim n_\chi^{\rm eq}(T)/s(T)\approx4.2\times10^{-3}\times(100/g_*)$, where $n_\chi^{\rm eq}(T)=3\zeta(3)T^3/(2\pi^2)$ is the equilibrium distribution before the FOPT. 
The mass is 
\begin{multline}
M_{\rm FB}\approx M_{\rm PBH}\approx 1.4\times10^{21}~{\rm g}\times v_b^3\left(\frac{\eta_\chi}{10^{-3}}\right)\times\\ 
\left(\frac{100}{g_*}\right)^{1/4}\left(\frac{100~{\rm GeV}}{T_*}\right)^2\left(\frac{100}{\beta/H}\right)^3\alpha^{1/4}~.
\end{multline}
Therefore our scenario prefers PBHs with sublunar mass or below. 
Denoting $R_{\rm Sch}^{}\equiv 2M_{\rm FB}/M_{\rm Pl}^2$ as the Schwarzschild radius, one obtains
\begin{multline}
\frac{R_{\rm FB}}{R_{\rm Sch}}\approx 2.3\times10^4\times v_b^{-2}\left(\frac{10^{-3}}{\eta_\chi}\right)^{2/3}\times\\
\left(\frac{100}{g_*}\right)^{1/6}\left(\frac{\beta/H}{100}\right)^2\alpha^{-1/2}~,
\end{multline}
which confirms that gravity is negligible in the Fermi-ball formation. 
On the other hand, for the formation of Fermi-balls, the initial range of the Yukawa force has to be smaller than $R_{\rm FB}/Q_{\rm FB}^{1/3}$, i.e. 
\be
L_\phi^{-1}|_{T_*}=\sqrt{\mu^2+c\,T_*^2}\gtrsim3.3\times g_\chi T_*\left(\frac{g_*}{100}\right)^{1/4}\alpha^{1/4}.
\ee
If the above condition is not satisfied, the old vacuum remnants directly collapse into PBHs instead of forming Fermi-balls as intermediate states. 
In that case, the PBH profile highly depends on the evolution trajectory of the remnants, and numerical simulation is necessary.

PBHs with mass between $10^9$~g and $10^{17}$~g evaporate between the Big Bang Nucleosynthesis (BBN) and today, leaving impacts on the BBN, Cosmic Microwave Background and extragalactic and Galactic $\gamma$-ray backgrounds, which in turn put stringent constraints on the PBH abundance. Such constraints are usually shown as upper limits for $\beta_{\rm PBH}^{'}$ as a function of PBH mass, where $\beta_{\rm PBH}^{'}$ is related to the energy fraction of PBH to the Universe at formation time~\cite{Carr:2020gox,Carr:2020xqk,Green:2020jor}. 
For our scenario,
\begin{multline}
\beta'_{\rm PBH}\approx 1.4\times10^{-15}\times v_b^{-3}\left(\frac{g_*}{100}\right)^{1/2}\times\\
\left(\frac{T_*}{100~{\rm GeV}}\right)^3\left(\frac{\beta/H}{100}\right)^3\left(\frac{M_{\rm PBH}}{10^{15}~{\rm g}}\right)^{3/2}~.
\end{multline}
For $M_{\rm PBH}\gtrsim10^9~{\rm g}$, the upper limit of $\beta'_{\rm PBH}$ varies from $10^{-29}$ to $10^{-17}$~\cite{Carr:2020gox,Carr:2020xqk,Green:2020jor}. For example, for $v_b=0.6$, $g_*=100$, $T_*=100$ GeV, and $\beta/H=100$, current bounds require $M_{\rm PBH}<10^{10}$~g, which corresponds to $\eta_\chi<1.8\times10^{-14}$ if $\alpha=1$.

PBHs with mass lager than $5.1\times10^{14}$~g can survive until today. 
Such PBHs can be probed/constrained by Hawking radiation (if $M_{\rm PBH}\lesssim10^{17}$ g), gravitational lensing, dynamical processes, cosmic structure, etc~\cite{Carr:2020gox,Carr:2020xqk,Green:2020jor}.  
The abundance is usually described by the fraction of PBH to DM, i.e. $f_{\rm PBH}=\Omega_{\rm PBH}/\Omega_{\rm DM}\leqslant1$. 
While there are already stringent constraints for $f_{\rm PBH}$, there is still a mass window $10^{17}~{\rm g}\sim10^{21}~{\rm g}$ that PBHs can account for all DM.\footnote{This window can be probed in the future~\cite{Jung:2019fcs,Laha:2019ssq,Dasgupta:2019cae,Laha:2020ivk,Ray:2021mxu}.}
In our scenario, we have
\begin{multline}
f_{\rm PBH}\approx
1.3\times 10^3\times v_b^{-3}\left(\frac{g_*}{100}\right)^{1/2}\left(\frac{T_*}{100~{\rm GeV}}\right)^3\times\\
\left(\frac{\beta/H}{100}\right)^3\left(\frac{M_{\rm PBH}}{10^{15}~{\rm g}
}\right)~,
\end{multline}
implying that non-evaporating PBHs can become a DM candidate when the formation temperature $T_*$ is significantly lower than $100$~GeV. For example, for $v_b=0.6$, $g_*=10$, $T_*=1.1$ GeV, $\alpha=1$, $\beta/H=100$ and $\eta_\chi=0.9\times10^{-10}$, one obtains $M_{\rm PBH}=4.2\times10^{17}$ g and $f_{\rm PBH}=1$, allowing a PBH DM candidate. 
Note that even in the mass region that PBHs cannot be the dominant DM component, they can still play an important role in the Universe evolution, such as seeding supermassive black holes or large scale structure formation. 

We emphasize that above discussions on $\beta_{\rm PBH}'$ and $f_{\rm PBH}$ apply only to an adiabatically expanding Universe. 
If there is some process happening between the formation of PBHs and the BBN that enhances the entropy of the Universe by a factor of $\Delta$, then the abundance is diluted as $\beta'_{\rm PBH}\to\beta'_{\rm PBH}/\Delta$ and $f_{\rm PBH}\to f_{\rm PBH}/\Delta$, and the constraints would be weakened. 
As there are various mechanisms generating significant entropy, such as thermal inflation~\cite{Lyth:1995hj,Lyth:1995ka,Asaka:1999xd}, early matter era~\cite{Scherrer:1984fd,Berlin:2016vnh,Berlin:2016gtr,Cosme:2020mck}, late-time decay of domain walls~\cite{Kawasaki:2004rx,Leite:2011sc,Hattori:2015xla}, etc,  
we can say that our PBH DM scenario is possible in many  particle physics models.

Finally, we comment on the wall velocity $v_b$. After nucleation, the bubble wall is accelerated for a short time, until the friction pressure $P$ from the plasma balances the vacuum pressure $\Delta U=U(0,T)-U(w(T),T)$, and then it reaches the ultimate velocity $v_b$. In principle, $v_b$ can be resolved by solving the Boltzmann equation provided that the particle content and interactions of the model are given. 
In our simple setup, only $\chi$ and $\bar{\chi}$ contribute to the pressure, 
and if we assume the thermal equilibrium of them and a 100\% reflection rate (i.e. $F_\chi^{\rm trap.}=1$), $P$ is given by~\cite{Chway:2019kft}
\be
P=\frac{7\pi^2}{180}\frac{(1+v_b)^3}{1-v_b^2}T^4~. 
\ee
For example, when $\Delta U=(100~{\rm GeV})^4$ and $T=80~{\rm GeV}$, the balancing condition $P=\Delta U$ yields $v_b=0.6$. For other values of $\Delta U$ and $T$, $v_b^{}$ typically becomes $0.2\sim0.8$. If there are other species contributing to the the friction pressure, $v_b$ can be reduced, which corresponds to the lighter PBHs, as $M_{\rm PBH}\propto v_b^3$.


\section{Summary}

We have discussed a novel PBH formation mechanism based on the collapse of Fermi-balls from a FOPT and illustrated that it can explain all of DM if $T_*^{}\ll 100$~GeV.     
When $T_*^{}\gtrsim 100$~GeV, our scenario typically predicts the overproduction of PBHs, and suitable dilution mechanism is needed to satisfy the experimental constraints. Since such a dilution is ubiquitous in new physics models, our scenario can be applicable in various models of particle physics.

Throughout the letter, we have assumed that a fermion $\chi$ has a vanishing bare mass and that the FOPT occurs by the transition of $\langle \phi\rangle$ from $\ave\phi=0$ to $w_*^{}$ for simplicity.  
Thus, $\chi$ is massless and massive in the false and true vacua, respectively. 
However, even if the fermion has a bare mass $M_0$ and the FOPT is generally from $\ave\phi=w_1^*$ to $w_2^*$, our mechanism still applies as long as $M_\chi^{}|_{\rm true}^{}=|M_0+g_\chi w_2^*|\gg M_\chi^{}|_{\rm false}^{}+T_*=|M_0+g_\chi w_1^*|+T_*$. 
In this case, the Fermi-ball/PBH profile needs to be modified, but the qualitative picture remains the same.

\section*{Acknowledgement}

We thank Sunghoon Jung, Taehun Kim, Alexander Kusenko, Osamu Seto and Yongcheng Wu for useful discussions. This work is supported by Grant Korea NRF-2019R1C1C1010050, and KPX also by the University of Nebraska-Lincoln.

\appendix
\section*{Appendix}

{\bf The Fermi-ball profile.}-- Given the charge $Q_{\rm FB}$, the energy of a Fermi-ball with radius $R$ at temperature $T$ is
\begin{multline}\label{FB_E_T}
E_{\rm FB}=\frac{3\pi}{4}\left(\frac{3}{2\pi}\right)^{2/3}\frac{Q_{\rm FB}^{4/3}}{R}\left[1+\frac{4\pi}{9}\left(\frac{2\pi}{3}\right)^{1/3}\frac{R^2T^2}{Q_{\rm FB}^{2/3}}\right]\\
+4\pi\sigma_0 R^2+\frac{4\pi}{3}U_0(T)R^3,
\end{multline}
where the first term is the Fermi gas kinetic energy derived from the low-temperature expansion of the Fermi integral, the second term is surface tension which is negligible since the Fermi-ball has a macroscopic size, and the third term is the bulk potential energy. By solving $dE_{\rm FB}/dR|_{R_{\rm FB}}=0$ one obtains the Fermi-ball profile
\bea\label{MR_T_profile}
M_{\rm FB}&=&Q_{\rm FB}\left(12\pi^2U_0(T)\right)^{1/4}\left(1+\frac{\pi\, T^2}{4\sqrt{3U_0(T)}}\right),\\
R_{\rm FB}&=&\left[\frac{3}{16}\left(\frac{3}{2\pi}\right)^{2/3}\frac{Q_{\rm FB}^{4/3}}{U_0(T)}\right]^{1/4}\left(1-\frac{\pi\, T^2}{12\sqrt{3U_0(T)}}\right).\nn
\eea
For a strong FOPT, $U_0^{1/4}(T)\gtrsim T$, the numbers in above brackets are very close to 1, leading to the expressions in Eq. (8) in the main text. The stable conditions for a Fermi-ball against decay and fission are respectively
\be
\frac{dM_{\rm FB}}{dQ_{\rm FB}}<g_\chi w(T),\quad \frac{d^2M_{\rm FB}}{dQ_{\rm FB}^2}<0;
\ee
While the second condition is automatically satisfied if the surface term ($\propto Q_{\rm FB}^{2/3}$) is considered, the first condition needs to be checked for a specific model.

{\bf Cooling of Fermi-balls.}-- When Fermi-balls are shrinking, the old vacuum remnant is in equilibrium via the fast $\chi\bar\chi\to\phi\phi$ process. 
After the Fermi-ball formation, pair annihilation stops since  only $\chi$'s are left inside Fermi-balls. 
Instead, the $\chi$ system cools down by emitting light fermions such as electrons or neutrinos~\cite{Witten:1984rs}.
Following the method in Ref.~\cite{Bai:2018dxf}, according to Stefan-Boltzmann's law, the energy escaping from a Fermi-ball per unit time is
\be
\mL(T)=\frac{N_f}{4}\times\frac{7}{8} \times \left(\frac{\pi^2}{30}T^4\right)\left(4\pi R_{\rm FB}^2\right)~,
\ee
where the $N_f$ is the number of DOF of light fermions, e.g. $6$ for $3$ generations of neutrinos. 
We then can establish the cooling equation for a Fermi-ball as
\be\label{cooling}
\frac{dM_{\rm FB}}{dt}=-\mL(T)~,
\ee
where $M_{\rm FB}$ is the Fermi-ball mass given in \Eq{MR_T_profile}. \Eq{cooling} can be solved analytically if we neglect the slow temperature dependence of $U_0(T)$.
Defining the cooling time scale $\tau_{\rm FB}$ as the time duration of a Fermi-ball cooling from $T$ to $T/2$, we find
\begin{multline}
\frac{\tau_{\rm FB}}{H^{-1}(T)}\approx0.16\times\left(\frac{3}{N_f}\right) v_b\left(\frac{g_*}{100}\right)^{7/12}\times\\
\left(\frac{\eta_\chi}{10^{-3}}\right)^{1/3}\left(\frac{100}{\beta/H}\right)\alpha^{1/4},
\end{multline}
which means $\tau_{\rm FB}$ is typically shorter than the Universe expansion time scale and Fermi-balls can cool down to $T_{\rm PBH}$ and collapse.

{\bf The Yukawa potential energy.}-- Assume a uniform distribution for $\chi$'s in the Fermi-ball, the full expression for Yukawa energy is
\be
\Delta E_{\rm FB}=-\frac{3g_\chi^2}{20\pi}\frac{Q_{\rm FB}^2}{R}f\left(\frac{L_\phi}{R}\right),
\ee
where
\be
f(\xi)=\frac{5}{2}\xi^2\left[1+\frac{3}{2}\xi\left(\xi^2-1\right)-\frac{3}{2}\xi(\xi +1)^2e^{-2/\xi}\right],
\ee
satisfying $f(0)=0$ and $f(\infty)=1$.

When $L_\phi/R\lesssim0.01$, $f(\xi)\approx5\xi^2/2$ is a very good approximation and the total Fermi-ball energy is
\be
E_{\rm FB}=\frac{3\pi}{4}\left(\frac{3}{2\pi}\right)^{2/3}\frac{Q_{\rm FB}^{4/3}}{R}-\frac{3g_\chi^2}{8\pi}\frac{Q_{\rm FB}^2L_\phi^2}{R^3}+\frac{4\pi}{3}U_0R^3.
\ee
To get the minimum of the energy, we take the first-order derivative, finding
\be
R^4\frac{dE_{\rm FB}}{dR}\propto u^3+p u+q,
\ee
where $u\equiv R^2$ and
\be
p=-\frac{3}{16}\left(\frac{3}{2\pi}\right)^{2/3}\frac{Q_{\rm FB}^{4/3}}{U_0},\quad q=\frac{9g_\chi^2}{32\pi^2}\frac{L_\phi^2Q_{\rm FB}^2}{U_0}.
\ee
Define the discriminant
\be\label{LphiU0}
\Delta=\left(\frac{q}{2}\right)^2+\left(\frac{p}{3}\right)^3\propto\frac{36}{\pi^2}g_\chi^4L_\phi^4U_0-1,
\ee
when $\Delta<0$, $u$ has two positive roots, with the larger one being the Fermi-ball radius square $R_{\rm FB}^2$, and the smaller one being the local maximum of the energy. If $\Delta\geqslant0$, then no Fermi-ball solution is found, and the old vacuum remnant just collapses to a PBH directly. This consideration already gives the condition for the Fermi-ball collapse,
\be
L_\phi>\sqrt{\frac{\pi}{6}}\frac{U_0^{-1/4}}{g_\chi}=\frac{1}{g_\chi}\sqrt{\frac{2\pi}{3\sqrt3}}\left(\frac{2\pi}{3}\right)^{1/6}\frac{R_{\rm FB}}{Q_{\rm FB}^{1/3}},
\ee
which is consistent with the naive estimate in the main text.

\bibliographystyle{apsrev}
\bibliography{reference}

\begin{thebibliography}{109}
\expandafter\ifx\csname natexlab\endcsname\relax\def\natexlab#1{#1}\fi
\expandafter\ifx\csname bibnamefont\endcsname\relax
  \def\bibnamefont#1{#1}\fi
\expandafter\ifx\csname bibfnamefont\endcsname\relax
  \def\bibfnamefont#1{#1}\fi
\expandafter\ifx\csname citenamefont\endcsname\relax
  \def\citenamefont#1{#1}\fi
\expandafter\ifx\csname url\endcsname\relax
  \def\url#1{\texttt{#1}}\fi
\expandafter\ifx\csname urlprefix\endcsname\relax\def\urlprefix{URL }\fi
\providecommand{\bibinfo}[2]{#2}
\providecommand{\eprint}[2][]{\url{#2}}

\bibitem[{\citenamefont{Zel'dovich and Novikov}(1967)}]{zel1967hypothesis}
\bibinfo{author}{\bibfnamefont{Y.~B.} \bibnamefont{Zel'dovich}}
  \bibnamefont{and} \bibinfo{author}{\bibfnamefont{I.}~\bibnamefont{Novikov}},
  \bibinfo{journal}{Soviet Astronomy} \textbf{\bibinfo{volume}{10}},
  \bibinfo{pages}{602} (\bibinfo{year}{1967}).

\bibitem[{\citenamefont{Hawking}(1971)}]{hawking1971gravitationally}
\bibinfo{author}{\bibfnamefont{S.}~\bibnamefont{Hawking}},
  \bibinfo{journal}{Monthly Notices of the Royal Astronomical Society}
  \textbf{\bibinfo{volume}{152}}, \bibinfo{pages}{75} (\bibinfo{year}{1971}).

\bibitem[{\citenamefont{Chapline}(1975)}]{Chapline:1975ojl}
\bibinfo{author}{\bibfnamefont{G.~F.} \bibnamefont{Chapline}},
  \bibinfo{journal}{Nature} \textbf{\bibinfo{volume}{253}},
  \bibinfo{pages}{251} (\bibinfo{year}{1975}).

\bibitem[{\citenamefont{Khlopov}(2010)}]{Khlopov:2008qy}
\bibinfo{author}{\bibfnamefont{M.~Y.} \bibnamefont{Khlopov}},
  \bibinfo{journal}{Res. Astron. Astrophys.} \textbf{\bibinfo{volume}{10}},
  \bibinfo{pages}{495} (\bibinfo{year}{2010}), \eprint{0801.0116}.

\bibitem[{\citenamefont{Carr et~al.}(2016)\citenamefont{Carr, Kuhnel, and
  Sandstad}}]{Carr:2016drx}
\bibinfo{author}{\bibfnamefont{B.}~\bibnamefont{Carr}},
  \bibinfo{author}{\bibfnamefont{F.}~\bibnamefont{Kuhnel}}, \bibnamefont{and}
  \bibinfo{author}{\bibfnamefont{M.}~\bibnamefont{Sandstad}},
  \bibinfo{journal}{Phys. Rev. D} \textbf{\bibinfo{volume}{94}},
  \bibinfo{pages}{083504} (\bibinfo{year}{2016}), \eprint{1607.06077}.

\bibitem[{\citenamefont{Carr et~al.}(2020)\citenamefont{Carr, Kohri, Sendouda,
  and Yokoyama}}]{Carr:2020gox}
\bibinfo{author}{\bibfnamefont{B.}~\bibnamefont{Carr}},
  \bibinfo{author}{\bibfnamefont{K.}~\bibnamefont{Kohri}},
  \bibinfo{author}{\bibfnamefont{Y.}~\bibnamefont{Sendouda}}, \bibnamefont{and}
  \bibinfo{author}{\bibfnamefont{J.}~\bibnamefont{Yokoyama}}
  (\bibinfo{year}{2020}), \eprint{2002.12778}.

\bibitem[{\citenamefont{Carr and Kuhnel}(2020)}]{Carr:2020xqk}
\bibinfo{author}{\bibfnamefont{B.}~\bibnamefont{Carr}} \bibnamefont{and}
  \bibinfo{author}{\bibfnamefont{F.}~\bibnamefont{Kuhnel}},
  \bibinfo{journal}{Ann. Rev. Nucl. Part. Sci.} \textbf{\bibinfo{volume}{70}},
  \bibinfo{pages}{355} (\bibinfo{year}{2020}), \eprint{2006.02838}.

\bibitem[{\citenamefont{Green and Kavanagh}(2021)}]{Green:2020jor}
\bibinfo{author}{\bibfnamefont{A.~M.} \bibnamefont{Green}} \bibnamefont{and}
  \bibinfo{author}{\bibfnamefont{B.~J.} \bibnamefont{Kavanagh}},
  \bibinfo{journal}{J. Phys. G} \textbf{\bibinfo{volume}{48}},
  \bibinfo{pages}{4} (\bibinfo{year}{2021}), \eprint{2007.10722}.

\bibitem[{\citenamefont{Bean and Magueijo}(2002)}]{Bean:2002kx}
\bibinfo{author}{\bibfnamefont{R.}~\bibnamefont{Bean}} \bibnamefont{and}
  \bibinfo{author}{\bibfnamefont{J.}~\bibnamefont{Magueijo}},
  \bibinfo{journal}{Phys. Rev. D} \textbf{\bibinfo{volume}{66}},
  \bibinfo{pages}{063505} (\bibinfo{year}{2002}), \eprint{astro-ph/0204486}.

\bibitem[{\citenamefont{Khlopov et~al.}(2005)\citenamefont{Khlopov, Rubin, and
  Sakharov}}]{Khlopov:2004sc}
\bibinfo{author}{\bibfnamefont{M.~Y.} \bibnamefont{Khlopov}},
  \bibinfo{author}{\bibfnamefont{S.~G.} \bibnamefont{Rubin}}, \bibnamefont{and}
  \bibinfo{author}{\bibfnamefont{A.~S.} \bibnamefont{Sakharov}},
  \bibinfo{journal}{Astropart. Phys.} \textbf{\bibinfo{volume}{23}},
  \bibinfo{pages}{265} (\bibinfo{year}{2005}), \eprint{astro-ph/0401532}.

\bibitem[{\citenamefont{Duechting}(2004)}]{Duechting:2004dk}
\bibinfo{author}{\bibfnamefont{N.}~\bibnamefont{Duechting}},
  \bibinfo{journal}{Phys. Rev. D} \textbf{\bibinfo{volume}{70}},
  \bibinfo{pages}{064015} (\bibinfo{year}{2004}), \eprint{astro-ph/0406260}.

\bibitem[{\citenamefont{Kawasaki et~al.}(2012)\citenamefont{Kawasaki, Kusenko,
  and Yanagida}}]{Kawasaki:2012kn}
\bibinfo{author}{\bibfnamefont{M.}~\bibnamefont{Kawasaki}},
  \bibinfo{author}{\bibfnamefont{A.}~\bibnamefont{Kusenko}}, \bibnamefont{and}
  \bibinfo{author}{\bibfnamefont{T.~T.} \bibnamefont{Yanagida}},
  \bibinfo{journal}{Phys. Lett. B} \textbf{\bibinfo{volume}{711}},
  \bibinfo{pages}{1} (\bibinfo{year}{2012}), \eprint{1202.3848}.

\bibitem[{\citenamefont{Clesse and Garc\'\i{}a-Bellido}(2015)}]{Clesse:2015wea}
\bibinfo{author}{\bibfnamefont{S.}~\bibnamefont{Clesse}} \bibnamefont{and}
  \bibinfo{author}{\bibfnamefont{J.}~\bibnamefont{Garc\'\i{}a-Bellido}},
  \bibinfo{journal}{Phys. Rev. D} \textbf{\bibinfo{volume}{92}},
  \bibinfo{pages}{023524} (\bibinfo{year}{2015}), \eprint{1501.07565}.

\bibitem[{\citenamefont{Hawking}(1974)}]{Hawking:1974rv}
\bibinfo{author}{\bibfnamefont{S.~W.} \bibnamefont{Hawking}},
  \bibinfo{journal}{Nature} \textbf{\bibinfo{volume}{248}}, \bibinfo{pages}{30}
  (\bibinfo{year}{1974}).

\bibitem[{\citenamefont{Carr}(1976)}]{carr1976some}
\bibinfo{author}{\bibfnamefont{B.~J.} \bibnamefont{Carr}},
  \bibinfo{journal}{The Astrophysical Journal} \textbf{\bibinfo{volume}{206}},
  \bibinfo{pages}{8} (\bibinfo{year}{1976}).

\bibitem[{\citenamefont{Baumann et~al.}(2007)\citenamefont{Baumann, Steinhardt,
  and Turok}}]{Baumann:2007yr}
\bibinfo{author}{\bibfnamefont{D.}~\bibnamefont{Baumann}},
  \bibinfo{author}{\bibfnamefont{P.~J.} \bibnamefont{Steinhardt}},
  \bibnamefont{and} \bibinfo{author}{\bibfnamefont{N.}~\bibnamefont{Turok}}
  (\bibinfo{year}{2007}), \eprint{hep-th/0703250}.

\bibitem[{\citenamefont{Hook}(2014)}]{Hook:2014mla}
\bibinfo{author}{\bibfnamefont{A.}~\bibnamefont{Hook}}, \bibinfo{journal}{Phys.
  Rev. D} \textbf{\bibinfo{volume}{90}}, \bibinfo{pages}{083535}
  (\bibinfo{year}{2014}), \eprint{1404.0113}.

\bibitem[{\citenamefont{Fujita et~al.}(2014)\citenamefont{Fujita, Kawasaki,
  Harigaya, and Matsuda}}]{Fujita:2014hha}
\bibinfo{author}{\bibfnamefont{T.}~\bibnamefont{Fujita}},
  \bibinfo{author}{\bibfnamefont{M.}~\bibnamefont{Kawasaki}},
  \bibinfo{author}{\bibfnamefont{K.}~\bibnamefont{Harigaya}}, \bibnamefont{and}
  \bibinfo{author}{\bibfnamefont{R.}~\bibnamefont{Matsuda}},
  \bibinfo{journal}{Phys. Rev. D} \textbf{\bibinfo{volume}{89}},
  \bibinfo{pages}{103501} (\bibinfo{year}{2014}), \eprint{1401.1909}.

\bibitem[{\citenamefont{Hamada and Iso}(2017)}]{Hamada:2016jnq}
\bibinfo{author}{\bibfnamefont{Y.}~\bibnamefont{Hamada}} \bibnamefont{and}
  \bibinfo{author}{\bibfnamefont{S.}~\bibnamefont{Iso}},
  \bibinfo{journal}{PTEP} \textbf{\bibinfo{volume}{2017}},
  \bibinfo{pages}{033B02} (\bibinfo{year}{2017}), \eprint{1610.02586}.

\bibitem[{\citenamefont{Morrison et~al.}(2019)\citenamefont{Morrison, Profumo,
  and Yu}}]{Morrison:2018xla}
\bibinfo{author}{\bibfnamefont{L.}~\bibnamefont{Morrison}},
  \bibinfo{author}{\bibfnamefont{S.}~\bibnamefont{Profumo}}, \bibnamefont{and}
  \bibinfo{author}{\bibfnamefont{Y.}~\bibnamefont{Yu}}, \bibinfo{journal}{JCAP}
  \textbf{\bibinfo{volume}{05}}, \bibinfo{pages}{005} (\bibinfo{year}{2019}),
  \eprint{1812.10606}.

\bibitem[{\citenamefont{Hooper and Krnjaic}(2021)}]{Hooper:2020otu}
\bibinfo{author}{\bibfnamefont{D.}~\bibnamefont{Hooper}} \bibnamefont{and}
  \bibinfo{author}{\bibfnamefont{G.}~\bibnamefont{Krnjaic}},
  \bibinfo{journal}{Phys. Rev. D} \textbf{\bibinfo{volume}{103}},
  \bibinfo{pages}{043504} (\bibinfo{year}{2021}), \eprint{2010.01134}.

\bibitem[{\citenamefont{Perez-Gonzalez and
  Turner}(2020)}]{Perez-Gonzalez:2020vnz}
\bibinfo{author}{\bibfnamefont{Y.~F.} \bibnamefont{Perez-Gonzalez}}
  \bibnamefont{and} \bibinfo{author}{\bibfnamefont{J.}~\bibnamefont{Turner}}
  (\bibinfo{year}{2020}), \eprint{2010.03565}.

\bibitem[{\citenamefont{Datta et~al.}(2020)\citenamefont{Datta, Ghosal, and
  Samanta}}]{Datta:2020bht}
\bibinfo{author}{\bibfnamefont{S.}~\bibnamefont{Datta}},
  \bibinfo{author}{\bibfnamefont{A.}~\bibnamefont{Ghosal}}, \bibnamefont{and}
  \bibinfo{author}{\bibfnamefont{R.}~\bibnamefont{Samanta}}
  (\bibinfo{year}{2020}), \eprint{2012.14981}.

\bibitem[{\citenamefont{Jyoti~Das et~al.}(2021)\citenamefont{Jyoti~Das,
  Mahanta, and Borah}}]{JyotiDas:2021shi}
\bibinfo{author}{\bibfnamefont{S.}~\bibnamefont{Jyoti~Das}},
  \bibinfo{author}{\bibfnamefont{D.}~\bibnamefont{Mahanta}}, \bibnamefont{and}
  \bibinfo{author}{\bibfnamefont{D.}~\bibnamefont{Borah}}
  (\bibinfo{year}{2021}), \eprint{2104.14496}.

\bibitem[{\citenamefont{Abbott et~al.}(2016{\natexlab{a}})}]{Abbott:2016blz}
\bibinfo{author}{\bibfnamefont{B.~P.} \bibnamefont{Abbott}}
  \bibnamefont{et~al.} (\bibinfo{collaboration}{LIGO Scientific, Virgo}),
  \bibinfo{journal}{Phys. Rev. Lett.} \textbf{\bibinfo{volume}{116}},
  \bibinfo{pages}{061102} (\bibinfo{year}{2016}{\natexlab{a}}),
  \eprint{1602.03837}.

\bibitem[{\citenamefont{Abbott et~al.}(2016{\natexlab{b}})}]{Abbott:2016nmj}
\bibinfo{author}{\bibfnamefont{B.~P.} \bibnamefont{Abbott}}
  \bibnamefont{et~al.} (\bibinfo{collaboration}{LIGO Scientific, Virgo}),
  \bibinfo{journal}{Phys. Rev. Lett.} \textbf{\bibinfo{volume}{116}},
  \bibinfo{pages}{241103} (\bibinfo{year}{2016}{\natexlab{b}}),
  \eprint{1606.04855}.

\bibitem[{\citenamefont{Abbott et~al.}(2017)}]{Abbott:2017vtc}
\bibinfo{author}{\bibfnamefont{B.~P.} \bibnamefont{Abbott}}
  \bibnamefont{et~al.} (\bibinfo{collaboration}{LIGO Scientific, VIRGO}),
  \bibinfo{journal}{Phys. Rev. Lett.} \textbf{\bibinfo{volume}{118}},
  \bibinfo{pages}{221101} (\bibinfo{year}{2017}), \bibinfo{note}{[Erratum:
  Phys.Rev.Lett. 121, 129901 (2018)]}, \eprint{1706.01812}.

\bibitem[{\citenamefont{Clesse and Garc\'\i{}a-Bellido}(2017)}]{Clesse:2016vqa}
\bibinfo{author}{\bibfnamefont{S.}~\bibnamefont{Clesse}} \bibnamefont{and}
  \bibinfo{author}{\bibfnamefont{J.}~\bibnamefont{Garc\'\i{}a-Bellido}},
  \bibinfo{journal}{Phys. Dark Univ.} \textbf{\bibinfo{volume}{15}},
  \bibinfo{pages}{142} (\bibinfo{year}{2017}), \eprint{1603.05234}.

\bibitem[{\citenamefont{Bird et~al.}(2016)\citenamefont{Bird, Cholis, Mu\~noz,
  Ali-Ha\"\i{}moud, Kamionkowski, Kovetz, Raccanelli, and
  Riess}}]{Bird:2016dcv}
\bibinfo{author}{\bibfnamefont{S.}~\bibnamefont{Bird}},
  \bibinfo{author}{\bibfnamefont{I.}~\bibnamefont{Cholis}},
  \bibinfo{author}{\bibfnamefont{J.~B.} \bibnamefont{Mu\~noz}},
  \bibinfo{author}{\bibfnamefont{Y.}~\bibnamefont{Ali-Ha\"\i{}moud}},
  \bibinfo{author}{\bibfnamefont{M.}~\bibnamefont{Kamionkowski}},
  \bibinfo{author}{\bibfnamefont{E.~D.} \bibnamefont{Kovetz}},
  \bibinfo{author}{\bibfnamefont{A.}~\bibnamefont{Raccanelli}},
  \bibnamefont{and} \bibinfo{author}{\bibfnamefont{A.~G.} \bibnamefont{Riess}},
  \bibinfo{journal}{Phys. Rev. Lett.} \textbf{\bibinfo{volume}{116}},
  \bibinfo{pages}{201301} (\bibinfo{year}{2016}), \eprint{1603.00464}.

\bibitem[{\citenamefont{Sasaki et~al.}(2016)\citenamefont{Sasaki, Suyama,
  Tanaka, and Yokoyama}}]{Sasaki:2016jop}
\bibinfo{author}{\bibfnamefont{M.}~\bibnamefont{Sasaki}},
  \bibinfo{author}{\bibfnamefont{T.}~\bibnamefont{Suyama}},
  \bibinfo{author}{\bibfnamefont{T.}~\bibnamefont{Tanaka}}, \bibnamefont{and}
  \bibinfo{author}{\bibfnamefont{S.}~\bibnamefont{Yokoyama}},
  \bibinfo{journal}{Phys. Rev. Lett.} \textbf{\bibinfo{volume}{117}},
  \bibinfo{pages}{061101} (\bibinfo{year}{2016}), \bibinfo{note}{[Erratum:
  Phys.Rev.Lett. 121, 059901 (2018)]}, \eprint{1603.08338}.

\bibitem[{\citenamefont{Carr and Hawking}(1974)}]{Carr:1974nx}
\bibinfo{author}{\bibfnamefont{B.~J.} \bibnamefont{Carr}} \bibnamefont{and}
  \bibinfo{author}{\bibfnamefont{S.~W.} \bibnamefont{Hawking}},
  \bibinfo{journal}{Mon. Not. Roy. Astron. Soc.}
  \textbf{\bibinfo{volume}{168}}, \bibinfo{pages}{399} (\bibinfo{year}{1974}).

\bibitem[{\citenamefont{Carr}(1975)}]{carr1975primordial}
\bibinfo{author}{\bibfnamefont{B.}~\bibnamefont{Carr}}, \bibinfo{journal}{The
  Astrophysical Journal} \textbf{\bibinfo{volume}{201}}, \bibinfo{pages}{1}
  (\bibinfo{year}{1975}).

\bibitem[{\citenamefont{Sasaki et~al.}(2018)\citenamefont{Sasaki, Suyama,
  Tanaka, and Yokoyama}}]{Sasaki:2018dmp}
\bibinfo{author}{\bibfnamefont{M.}~\bibnamefont{Sasaki}},
  \bibinfo{author}{\bibfnamefont{T.}~\bibnamefont{Suyama}},
  \bibinfo{author}{\bibfnamefont{T.}~\bibnamefont{Tanaka}}, \bibnamefont{and}
  \bibinfo{author}{\bibfnamefont{S.}~\bibnamefont{Yokoyama}},
  \bibinfo{journal}{Class. Quant. Grav.} \textbf{\bibinfo{volume}{35}},
  \bibinfo{pages}{063001} (\bibinfo{year}{2018}), \eprint{1801.05235}.

\bibitem[{\citenamefont{Hawking}(1989)}]{Hawking:1987bn}
\bibinfo{author}{\bibfnamefont{S.~W.} \bibnamefont{Hawking}},
  \bibinfo{journal}{Phys. Lett. B} \textbf{\bibinfo{volume}{231}},
  \bibinfo{pages}{237} (\bibinfo{year}{1989}).

\bibitem[{\citenamefont{Caldwell and Casper}(1996)}]{Caldwell:1995fu}
\bibinfo{author}{\bibfnamefont{R.~R.} \bibnamefont{Caldwell}} \bibnamefont{and}
  \bibinfo{author}{\bibfnamefont{P.}~\bibnamefont{Casper}},
  \bibinfo{journal}{Phys. Rev. D} \textbf{\bibinfo{volume}{53}},
  \bibinfo{pages}{3002} (\bibinfo{year}{1996}), \eprint{gr-qc/9509012}.

\bibitem[{\citenamefont{Garriga and Vilenkin}(1993)}]{Garriga:1992nm}
\bibinfo{author}{\bibfnamefont{J.}~\bibnamefont{Garriga}} \bibnamefont{and}
  \bibinfo{author}{\bibfnamefont{A.}~\bibnamefont{Vilenkin}},
  \bibinfo{journal}{Phys. Rev. D} \textbf{\bibinfo{volume}{47}},
  \bibinfo{pages}{3265} (\bibinfo{year}{1993}), \eprint{hep-ph/9208212}.

\bibitem[{\citenamefont{Rubin et~al.}(2000)\citenamefont{Rubin, Khlopov, and
  Sakharov}}]{Rubin:2000dq}
\bibinfo{author}{\bibfnamefont{S.~G.} \bibnamefont{Rubin}},
  \bibinfo{author}{\bibfnamefont{M.~Y.} \bibnamefont{Khlopov}},
  \bibnamefont{and} \bibinfo{author}{\bibfnamefont{A.~S.}
  \bibnamefont{Sakharov}}, \bibinfo{journal}{Grav. Cosmol.}
  \textbf{\bibinfo{volume}{6}}, \bibinfo{pages}{51} (\bibinfo{year}{2000}),
  \eprint{hep-ph/0005271}.

\bibitem[{\citenamefont{Rubin et~al.}(2001)\citenamefont{Rubin, Sakharov, and
  Khlopov}}]{Rubin:2001yw}
\bibinfo{author}{\bibfnamefont{S.~G.} \bibnamefont{Rubin}},
  \bibinfo{author}{\bibfnamefont{A.~S.} \bibnamefont{Sakharov}},
  \bibnamefont{and} \bibinfo{author}{\bibfnamefont{M.~Y.}
  \bibnamefont{Khlopov}}, \bibinfo{journal}{J. Exp. Theor. Phys.}
  \textbf{\bibinfo{volume}{91}}, \bibinfo{pages}{921} (\bibinfo{year}{2001}),
  \eprint{hep-ph/0106187}.

\bibitem[{\citenamefont{Dokuchaev et~al.}(2005)\citenamefont{Dokuchaev,
  Eroshenko, and Rubin}}]{Dokuchaev:2004kr}
\bibinfo{author}{\bibfnamefont{V.}~\bibnamefont{Dokuchaev}},
  \bibinfo{author}{\bibfnamefont{Y.}~\bibnamefont{Eroshenko}},
  \bibnamefont{and} \bibinfo{author}{\bibfnamefont{S.}~\bibnamefont{Rubin}},
  \bibinfo{journal}{Grav. Cosmol.} \textbf{\bibinfo{volume}{11}},
  \bibinfo{pages}{99} (\bibinfo{year}{2005}), \eprint{astro-ph/0412418}.

\bibitem[{\citenamefont{Deng et~al.}(2017)\citenamefont{Deng, Garriga, and
  Vilenkin}}]{Deng:2016vzb}
\bibinfo{author}{\bibfnamefont{H.}~\bibnamefont{Deng}},
  \bibinfo{author}{\bibfnamefont{J.}~\bibnamefont{Garriga}}, \bibnamefont{and}
  \bibinfo{author}{\bibfnamefont{A.}~\bibnamefont{Vilenkin}},
  \bibinfo{journal}{JCAP} \textbf{\bibinfo{volume}{04}}, \bibinfo{pages}{050}
  (\bibinfo{year}{2017}), \eprint{1612.03753}.

\bibitem[{\citenamefont{Cotner and
  Kusenko}(2017{\natexlab{a}})}]{Cotner:2016cvr}
\bibinfo{author}{\bibfnamefont{E.}~\bibnamefont{Cotner}} \bibnamefont{and}
  \bibinfo{author}{\bibfnamefont{A.}~\bibnamefont{Kusenko}},
  \bibinfo{journal}{Phys. Rev. Lett.} \textbf{\bibinfo{volume}{119}},
  \bibinfo{pages}{031103} (\bibinfo{year}{2017}{\natexlab{a}}),
  \eprint{1612.02529}.

\bibitem[{\citenamefont{Cotner and
  Kusenko}(2017{\natexlab{b}})}]{Cotner:2017tir}
\bibinfo{author}{\bibfnamefont{E.}~\bibnamefont{Cotner}} \bibnamefont{and}
  \bibinfo{author}{\bibfnamefont{A.}~\bibnamefont{Kusenko}},
  \bibinfo{journal}{Phys. Rev. D} \textbf{\bibinfo{volume}{96}},
  \bibinfo{pages}{103002} (\bibinfo{year}{2017}{\natexlab{b}}),
  \eprint{1706.09003}.

\bibitem[{\citenamefont{Cotner et~al.}(2018)\citenamefont{Cotner, Kusenko, and
  Takhistov}}]{Cotner:2018vug}
\bibinfo{author}{\bibfnamefont{E.}~\bibnamefont{Cotner}},
  \bibinfo{author}{\bibfnamefont{A.}~\bibnamefont{Kusenko}}, \bibnamefont{and}
  \bibinfo{author}{\bibfnamefont{V.}~\bibnamefont{Takhistov}},
  \bibinfo{journal}{Phys. Rev. D} \textbf{\bibinfo{volume}{98}},
  \bibinfo{pages}{083513} (\bibinfo{year}{2018}), \eprint{1801.03321}.

\bibitem[{\citenamefont{Cotner et~al.}(2019)\citenamefont{Cotner, Kusenko,
  Sasaki, and Takhistov}}]{Cotner:2019ykd}
\bibinfo{author}{\bibfnamefont{E.}~\bibnamefont{Cotner}},
  \bibinfo{author}{\bibfnamefont{A.}~\bibnamefont{Kusenko}},
  \bibinfo{author}{\bibfnamefont{M.}~\bibnamefont{Sasaki}}, \bibnamefont{and}
  \bibinfo{author}{\bibfnamefont{V.}~\bibnamefont{Takhistov}},
  \bibinfo{journal}{JCAP} \textbf{\bibinfo{volume}{10}}, \bibinfo{pages}{077}
  (\bibinfo{year}{2019}), \eprint{1907.10613}.

\bibitem[{\citenamefont{Crawford and Schramm}(1982)}]{Crawford:1982yz}
\bibinfo{author}{\bibfnamefont{M.}~\bibnamefont{Crawford}} \bibnamefont{and}
  \bibinfo{author}{\bibfnamefont{D.~N.} \bibnamefont{Schramm}},
  \bibinfo{journal}{Nature} \textbf{\bibinfo{volume}{298}},
  \bibinfo{pages}{538} (\bibinfo{year}{1982}).

\bibitem[{\citenamefont{Hawking et~al.}(1982)\citenamefont{Hawking, Moss, and
  Stewart}}]{Hawking:1982ga}
\bibinfo{author}{\bibfnamefont{S.~W.} \bibnamefont{Hawking}},
  \bibinfo{author}{\bibfnamefont{I.~G.} \bibnamefont{Moss}}, \bibnamefont{and}
  \bibinfo{author}{\bibfnamefont{J.~M.} \bibnamefont{Stewart}},
  \bibinfo{journal}{Phys. Rev. D} \textbf{\bibinfo{volume}{26}},
  \bibinfo{pages}{2681} (\bibinfo{year}{1982}).

\bibitem[{\citenamefont{La and Steinhardt}(1989)}]{La:1989st}
\bibinfo{author}{\bibfnamefont{D.}~\bibnamefont{La}} \bibnamefont{and}
  \bibinfo{author}{\bibfnamefont{P.~J.} \bibnamefont{Steinhardt}},
  \bibinfo{journal}{Phys. Lett. B} \textbf{\bibinfo{volume}{220}},
  \bibinfo{pages}{375} (\bibinfo{year}{1989}).

\bibitem[{\citenamefont{Moss}(1994)}]{Moss:1994iq}
\bibinfo{author}{\bibfnamefont{I.~G.} \bibnamefont{Moss}},
  \bibinfo{journal}{Phys. Rev. D} \textbf{\bibinfo{volume}{50}},
  \bibinfo{pages}{676} (\bibinfo{year}{1994}).

\bibitem[{\citenamefont{Konoplich et~al.}(1998)\citenamefont{Konoplich, Rubin,
  Sakharov, and Khlopov}}]{konoplich1998formation}
\bibinfo{author}{\bibfnamefont{R.}~\bibnamefont{Konoplich}},
  \bibinfo{author}{\bibfnamefont{S.}~\bibnamefont{Rubin}},
  \bibinfo{author}{\bibfnamefont{A.}~\bibnamefont{Sakharov}}, \bibnamefont{and}
  \bibinfo{author}{\bibfnamefont{M.~Y.} \bibnamefont{Khlopov}},
  \bibinfo{journal}{Astronomy Letters} \textbf{\bibinfo{volume}{24}},
  \bibinfo{pages}{413} (\bibinfo{year}{1998}).

\bibitem[{\citenamefont{Konoplich et~al.}(1999)\citenamefont{Konoplich, Rubin,
  Sakharov, and Khlopov}}]{Konoplich:1999qq}
\bibinfo{author}{\bibfnamefont{R.~V.} \bibnamefont{Konoplich}},
  \bibinfo{author}{\bibfnamefont{S.~G.} \bibnamefont{Rubin}},
  \bibinfo{author}{\bibfnamefont{A.~S.} \bibnamefont{Sakharov}},
  \bibnamefont{and} \bibinfo{author}{\bibfnamefont{M.~Y.}
  \bibnamefont{Khlopov}}, \bibinfo{journal}{Phys. Atom. Nucl.}
  \textbf{\bibinfo{volume}{62}}, \bibinfo{pages}{1593} (\bibinfo{year}{1999}).

\bibitem[{\citenamefont{Kodama et~al.}(1982)\citenamefont{Kodama, Sasaki, and
  Sato}}]{Kodama:1982sf}
\bibinfo{author}{\bibfnamefont{H.}~\bibnamefont{Kodama}},
  \bibinfo{author}{\bibfnamefont{M.}~\bibnamefont{Sasaki}}, \bibnamefont{and}
  \bibinfo{author}{\bibfnamefont{K.}~\bibnamefont{Sato}},
  \bibinfo{journal}{Prog. Theor. Phys.} \textbf{\bibinfo{volume}{68}},
  \bibinfo{pages}{1979} (\bibinfo{year}{1982}).

\bibitem[{\citenamefont{Kusenko et~al.}(2020)\citenamefont{Kusenko, Sasaki,
  Sugiyama, Takada, Takhistov, and Vitagliano}}]{Kusenko:2020pcg}
\bibinfo{author}{\bibfnamefont{A.}~\bibnamefont{Kusenko}},
  \bibinfo{author}{\bibfnamefont{M.}~\bibnamefont{Sasaki}},
  \bibinfo{author}{\bibfnamefont{S.}~\bibnamefont{Sugiyama}},
  \bibinfo{author}{\bibfnamefont{M.}~\bibnamefont{Takada}},
  \bibinfo{author}{\bibfnamefont{V.}~\bibnamefont{Takhistov}},
  \bibnamefont{and}
  \bibinfo{author}{\bibfnamefont{E.}~\bibnamefont{Vitagliano}},
  \bibinfo{journal}{Phys. Rev. Lett.} \textbf{\bibinfo{volume}{125}},
  \bibinfo{pages}{181304} (\bibinfo{year}{2020}), \eprint{2001.09160}.

\bibitem[{\citenamefont{Baker et~al.}(2021)\citenamefont{Baker, Breitbach,
  Kopp, and Mittnacht}}]{Baker:2021nyl}
\bibinfo{author}{\bibfnamefont{M.~J.} \bibnamefont{Baker}},
  \bibinfo{author}{\bibfnamefont{M.}~\bibnamefont{Breitbach}},
  \bibinfo{author}{\bibfnamefont{J.}~\bibnamefont{Kopp}}, \bibnamefont{and}
  \bibinfo{author}{\bibfnamefont{L.}~\bibnamefont{Mittnacht}}
  (\bibinfo{year}{2021}), \eprint{2105.07481}.

\bibitem[{\citenamefont{Hong et~al.}(2020)\citenamefont{Hong, Jung, and
  Xie}}]{Hong:2020est}
\bibinfo{author}{\bibfnamefont{J.-P.} \bibnamefont{Hong}},
  \bibinfo{author}{\bibfnamefont{S.}~\bibnamefont{Jung}}, \bibnamefont{and}
  \bibinfo{author}{\bibfnamefont{K.-P.} \bibnamefont{Xie}},
  \bibinfo{journal}{Phys. Rev. D} \textbf{\bibinfo{volume}{102}},
  \bibinfo{pages}{075028} (\bibinfo{year}{2020}), \eprint{2008.04430}.

\bibitem[{\citenamefont{Creminelli et~al.}(2002)\citenamefont{Creminelli,
  Nicolis, and Rattazzi}}]{Creminelli:2001th}
\bibinfo{author}{\bibfnamefont{P.}~\bibnamefont{Creminelli}},
  \bibinfo{author}{\bibfnamefont{A.}~\bibnamefont{Nicolis}}, \bibnamefont{and}
  \bibinfo{author}{\bibfnamefont{R.}~\bibnamefont{Rattazzi}},
  \bibinfo{journal}{JHEP} \textbf{\bibinfo{volume}{03}}, \bibinfo{pages}{051}
  (\bibinfo{year}{2002}), \eprint{hep-th/0107141}.

\bibitem[{\citenamefont{Nardini et~al.}(2007)\citenamefont{Nardini, Quiros, and
  Wulzer}}]{Nardini:2007me}
\bibinfo{author}{\bibfnamefont{G.}~\bibnamefont{Nardini}},
  \bibinfo{author}{\bibfnamefont{M.}~\bibnamefont{Quiros}}, \bibnamefont{and}
  \bibinfo{author}{\bibfnamefont{A.}~\bibnamefont{Wulzer}},
  \bibinfo{journal}{JHEP} \textbf{\bibinfo{volume}{09}}, \bibinfo{pages}{077}
  (\bibinfo{year}{2007}), \eprint{0706.3388}.

\bibitem[{\citenamefont{Konstandin and Servant}(2011)}]{Konstandin:2011dr}
\bibinfo{author}{\bibfnamefont{T.}~\bibnamefont{Konstandin}} \bibnamefont{and}
  \bibinfo{author}{\bibfnamefont{G.}~\bibnamefont{Servant}},
  \bibinfo{journal}{JCAP} \textbf{\bibinfo{volume}{12}}, \bibinfo{pages}{009}
  (\bibinfo{year}{2011}), \eprint{1104.4791}.

\bibitem[{\citenamefont{Jinno and Takimoto}(2017)}]{Jinno:2016knw}
\bibinfo{author}{\bibfnamefont{R.}~\bibnamefont{Jinno}} \bibnamefont{and}
  \bibinfo{author}{\bibfnamefont{M.}~\bibnamefont{Takimoto}},
  \bibinfo{journal}{Phys. Rev. D} \textbf{\bibinfo{volume}{95}},
  \bibinfo{pages}{015020} (\bibinfo{year}{2017}), \eprint{1604.05035}.

\bibitem[{\citenamefont{Marzo et~al.}(2019)\citenamefont{Marzo, Marzola, and
  Vaskonen}}]{Marzo:2018nov}
\bibinfo{author}{\bibfnamefont{C.}~\bibnamefont{Marzo}},
  \bibinfo{author}{\bibfnamefont{L.}~\bibnamefont{Marzola}}, \bibnamefont{and}
  \bibinfo{author}{\bibfnamefont{V.}~\bibnamefont{Vaskonen}},
  \bibinfo{journal}{Eur. Phys. J. C} \textbf{\bibinfo{volume}{79}},
  \bibinfo{pages}{601} (\bibinfo{year}{2019}), \eprint{1811.11169}.

\bibitem[{\citenamefont{Hambye et~al.}(2018)\citenamefont{Hambye, Strumia, and
  Teresi}}]{Hambye:2018qjv}
\bibinfo{author}{\bibfnamefont{T.}~\bibnamefont{Hambye}},
  \bibinfo{author}{\bibfnamefont{A.}~\bibnamefont{Strumia}}, \bibnamefont{and}
  \bibinfo{author}{\bibfnamefont{D.}~\bibnamefont{Teresi}},
  \bibinfo{journal}{JHEP} \textbf{\bibinfo{volume}{08}}, \bibinfo{pages}{188}
  (\bibinfo{year}{2018}), \eprint{1805.01473}.

\bibitem[{\citenamefont{Baratella et~al.}(2019)\citenamefont{Baratella,
  Pomarol, and Rompineve}}]{Baratella:2018pxi}
\bibinfo{author}{\bibfnamefont{P.}~\bibnamefont{Baratella}},
  \bibinfo{author}{\bibfnamefont{A.}~\bibnamefont{Pomarol}}, \bibnamefont{and}
  \bibinfo{author}{\bibfnamefont{F.}~\bibnamefont{Rompineve}},
  \bibinfo{journal}{JHEP} \textbf{\bibinfo{volume}{03}}, \bibinfo{pages}{100}
  (\bibinfo{year}{2019}), \eprint{1812.06996}.

\bibitem[{\citenamefont{Carena et~al.}(2005)\citenamefont{Carena, Megevand,
  Quiros, and Wagner}}]{Carena:2004ha}
\bibinfo{author}{\bibfnamefont{M.}~\bibnamefont{Carena}},
  \bibinfo{author}{\bibfnamefont{A.}~\bibnamefont{Megevand}},
  \bibinfo{author}{\bibfnamefont{M.}~\bibnamefont{Quiros}}, \bibnamefont{and}
  \bibinfo{author}{\bibfnamefont{C.~E.} \bibnamefont{Wagner}},
  \bibinfo{journal}{Nucl. Phys. B} \textbf{\bibinfo{volume}{716}},
  \bibinfo{pages}{319} (\bibinfo{year}{2005}), \eprint{hep-ph/0410352}.

\bibitem[{\citenamefont{Angelescu and Huang}(2019)}]{Angelescu:2018dkk}
\bibinfo{author}{\bibfnamefont{A.}~\bibnamefont{Angelescu}} \bibnamefont{and}
  \bibinfo{author}{\bibfnamefont{P.}~\bibnamefont{Huang}},
  \bibinfo{journal}{Phys. Rev. D} \textbf{\bibinfo{volume}{99}},
  \bibinfo{pages}{055023} (\bibinfo{year}{2019}), \eprint{1812.08293}.

\bibitem[{\citenamefont{Kaplan et~al.}(2009)\citenamefont{Kaplan, Luty, and
  Zurek}}]{Kaplan:2009ag}
\bibinfo{author}{\bibfnamefont{D.~E.} \bibnamefont{Kaplan}},
  \bibinfo{author}{\bibfnamefont{M.~A.} \bibnamefont{Luty}}, \bibnamefont{and}
  \bibinfo{author}{\bibfnamefont{K.~M.} \bibnamefont{Zurek}},
  \bibinfo{journal}{Phys. Rev. D} \textbf{\bibinfo{volume}{79}},
  \bibinfo{pages}{115016} (\bibinfo{year}{2009}), \eprint{0901.4117}.

\bibitem[{\citenamefont{Petraki and Volkas}(2013)}]{Petraki:2013wwa}
\bibinfo{author}{\bibfnamefont{K.}~\bibnamefont{Petraki}} \bibnamefont{and}
  \bibinfo{author}{\bibfnamefont{R.~R.} \bibnamefont{Volkas}},
  \bibinfo{journal}{Int. J. Mod. Phys. A} \textbf{\bibinfo{volume}{28}},
  \bibinfo{pages}{1330028} (\bibinfo{year}{2013}), \eprint{1305.4939}.

\bibitem[{\citenamefont{Zurek}(2014)}]{Zurek:2013wia}
\bibinfo{author}{\bibfnamefont{K.~M.} \bibnamefont{Zurek}},
  \bibinfo{journal}{Phys. Rept.} \textbf{\bibinfo{volume}{537}},
  \bibinfo{pages}{91} (\bibinfo{year}{2014}), \eprint{1308.0338}.

\bibitem[{\citenamefont{Gross et~al.}(2021)\citenamefont{Gross, Landini,
  Strumia, and Teresi}}]{Gross:2021qgx}
\bibinfo{author}{\bibfnamefont{C.}~\bibnamefont{Gross}},
  \bibinfo{author}{\bibfnamefont{G.}~\bibnamefont{Landini}},
  \bibinfo{author}{\bibfnamefont{A.}~\bibnamefont{Strumia}}, \bibnamefont{and}
  \bibinfo{author}{\bibfnamefont{D.}~\bibnamefont{Teresi}}
  (\bibinfo{year}{2021}), \eprint{2105.02840}.

\bibitem[{\citenamefont{Amendola et~al.}(2018)\citenamefont{Amendola, Rubio,
  and Wetterich}}]{Amendola:2017xhl}
\bibinfo{author}{\bibfnamefont{L.}~\bibnamefont{Amendola}},
  \bibinfo{author}{\bibfnamefont{J.}~\bibnamefont{Rubio}}, \bibnamefont{and}
  \bibinfo{author}{\bibfnamefont{C.}~\bibnamefont{Wetterich}},
  \bibinfo{journal}{Phys. Rev. D} \textbf{\bibinfo{volume}{97}},
  \bibinfo{pages}{081302} (\bibinfo{year}{2018}), \eprint{1711.09915}.

\bibitem[{\citenamefont{Flores and Kusenko}(2021)}]{Flores:2020drq}
\bibinfo{author}{\bibfnamefont{M.~M.} \bibnamefont{Flores}} \bibnamefont{and}
  \bibinfo{author}{\bibfnamefont{A.}~\bibnamefont{Kusenko}},
  \bibinfo{journal}{Phys. Rev. Lett.} \textbf{\bibinfo{volume}{126}},
  \bibinfo{pages}{041101} (\bibinfo{year}{2021}), \eprint{2008.12456}.

\bibitem[{\citenamefont{Linde}(1983)}]{Linde:1981zj}
\bibinfo{author}{\bibfnamefont{A.~D.} \bibnamefont{Linde}},
  \bibinfo{journal}{Nucl. Phys. B} \textbf{\bibinfo{volume}{216}},
  \bibinfo{pages}{421} (\bibinfo{year}{1983}), \bibinfo{note}{[Erratum:
  Nucl.Phys.B 223, 544 (1983)]}.

\bibitem[{\citenamefont{Guth and Weinberg}(1981)}]{Guth:1981uk}
\bibinfo{author}{\bibfnamefont{A.~H.} \bibnamefont{Guth}} \bibnamefont{and}
  \bibinfo{author}{\bibfnamefont{E.~J.} \bibnamefont{Weinberg}},
  \bibinfo{journal}{Phys. Rev. D} \textbf{\bibinfo{volume}{23}},
  \bibinfo{pages}{876} (\bibinfo{year}{1981}).

\bibitem[{\citenamefont{Ellis et~al.}(2019)\citenamefont{Ellis, Lewicki, No,
  and Vaskonen}}]{Ellis:2019oqb}
\bibinfo{author}{\bibfnamefont{J.}~\bibnamefont{Ellis}},
  \bibinfo{author}{\bibfnamefont{M.}~\bibnamefont{Lewicki}},
  \bibinfo{author}{\bibfnamefont{J.~M.} \bibnamefont{No}}, \bibnamefont{and}
  \bibinfo{author}{\bibfnamefont{V.}~\bibnamefont{Vaskonen}},
  \bibinfo{journal}{JCAP} \textbf{\bibinfo{volume}{06}}, \bibinfo{pages}{024}
  (\bibinfo{year}{2019}), \eprint{1903.09642}.

\bibitem[{\citenamefont{Wang et~al.}(2020)\citenamefont{Wang, Huang, and
  Zhang}}]{Wang:2020jrd}
\bibinfo{author}{\bibfnamefont{X.}~\bibnamefont{Wang}},
  \bibinfo{author}{\bibfnamefont{F.~P.} \bibnamefont{Huang}}, \bibnamefont{and}
  \bibinfo{author}{\bibfnamefont{X.}~\bibnamefont{Zhang}},
  \bibinfo{journal}{JCAP} \textbf{\bibinfo{volume}{05}}, \bibinfo{pages}{045}
  (\bibinfo{year}{2020}), \eprint{2003.08892}.

\bibitem[{\citenamefont{Rintoul and Torquato}(1997)}]{rintoul1997precise}
\bibinfo{author}{\bibfnamefont{M.~D.} \bibnamefont{Rintoul}} \bibnamefont{and}
  \bibinfo{author}{\bibfnamefont{S.}~\bibnamefont{Torquato}},
  \bibinfo{journal}{Journal of physics a: mathematical and general}
  \textbf{\bibinfo{volume}{30}}, \bibinfo{pages}{L585} (\bibinfo{year}{1997}).

\bibitem[{\citenamefont{Chway et~al.}(2020)\citenamefont{Chway, Jung, and
  Shin}}]{Chway:2019kft}
\bibinfo{author}{\bibfnamefont{D.}~\bibnamefont{Chway}},
  \bibinfo{author}{\bibfnamefont{T.~H.} \bibnamefont{Jung}}, \bibnamefont{and}
  \bibinfo{author}{\bibfnamefont{C.~S.} \bibnamefont{Shin}},
  \bibinfo{journal}{Phys. Rev. D} \textbf{\bibinfo{volume}{101}},
  \bibinfo{pages}{095019} (\bibinfo{year}{2020}), \eprint{1912.04238}.

\bibitem[{\citenamefont{Macpherson and Campbell}(1995)}]{Macpherson:1994wf}
\bibinfo{author}{\bibfnamefont{A.~L.} \bibnamefont{Macpherson}}
  \bibnamefont{and} \bibinfo{author}{\bibfnamefont{B.~A.}
  \bibnamefont{Campbell}}, \bibinfo{journal}{Phys. Lett. B}
  \textbf{\bibinfo{volume}{347}}, \bibinfo{pages}{205} (\bibinfo{year}{1995}),
  \eprint{hep-ph/9408387}.

\bibitem[{\citenamefont{Macpherson and Pinfold}(1994)}]{Macpherson:1994vk}
\bibinfo{author}{\bibfnamefont{A.~L.} \bibnamefont{Macpherson}}
  \bibnamefont{and} \bibinfo{author}{\bibfnamefont{J.~L.}
  \bibnamefont{Pinfold}} (\bibinfo{year}{1994}), \eprint{hep-ph/9412264}.

\bibitem[{\citenamefont{Sivaram and Arun}(2011)}]{Sivaram:2011jk}
\bibinfo{author}{\bibfnamefont{C.}~\bibnamefont{Sivaram}} \bibnamefont{and}
  \bibinfo{author}{\bibfnamefont{K.}~\bibnamefont{Arun}}
  (\bibinfo{year}{2011}), \eprint{1109.5266}.

\bibitem[{\citenamefont{Witten}(1984)}]{Witten:1984rs}
\bibinfo{author}{\bibfnamefont{E.}~\bibnamefont{Witten}},
  \bibinfo{journal}{Phys. Rev. D} \textbf{\bibinfo{volume}{30}},
  \bibinfo{pages}{272} (\bibinfo{year}{1984}).

\bibitem[{\citenamefont{Frieman and Giudice}(1991)}]{Frieman:1990nh}
\bibinfo{author}{\bibfnamefont{J.~A.} \bibnamefont{Frieman}} \bibnamefont{and}
  \bibinfo{author}{\bibfnamefont{G.~F.} \bibnamefont{Giudice}},
  \bibinfo{journal}{Nucl. Phys. B} \textbf{\bibinfo{volume}{355}},
  \bibinfo{pages}{162} (\bibinfo{year}{1991}).

\bibitem[{\citenamefont{Zhitnitsky}(2003)}]{Zhitnitsky:2002qa}
\bibinfo{author}{\bibfnamefont{A.~R.} \bibnamefont{Zhitnitsky}},
  \bibinfo{journal}{JCAP} \textbf{\bibinfo{volume}{10}}, \bibinfo{pages}{010}
  (\bibinfo{year}{2003}), \eprint{hep-ph/0202161}.

\bibitem[{\citenamefont{Oaknin and Zhitnitsky}(2005)}]{Oaknin:2003uv}
\bibinfo{author}{\bibfnamefont{D.~H.} \bibnamefont{Oaknin}} \bibnamefont{and}
  \bibinfo{author}{\bibfnamefont{A.}~\bibnamefont{Zhitnitsky}},
  \bibinfo{journal}{Phys. Rev. D} \textbf{\bibinfo{volume}{71}},
  \bibinfo{pages}{023519} (\bibinfo{year}{2005}), \eprint{hep-ph/0309086}.

\bibitem[{\citenamefont{Lawson and Zhitnitsky}(2013)}]{Lawson:2012zu}
\bibinfo{author}{\bibfnamefont{K.}~\bibnamefont{Lawson}} \bibnamefont{and}
  \bibinfo{author}{\bibfnamefont{A.~R.} \bibnamefont{Zhitnitsky}},
  \bibinfo{journal}{Phys. Lett. B} \textbf{\bibinfo{volume}{724}},
  \bibinfo{pages}{17} (\bibinfo{year}{2013}), \eprint{1210.2400}.

\bibitem[{\citenamefont{Atreya et~al.}(2014)\citenamefont{Atreya, Sarkar, and
  Srivastava}}]{Atreya:2014sca}
\bibinfo{author}{\bibfnamefont{A.}~\bibnamefont{Atreya}},
  \bibinfo{author}{\bibfnamefont{A.}~\bibnamefont{Sarkar}}, \bibnamefont{and}
  \bibinfo{author}{\bibfnamefont{A.~M.} \bibnamefont{Srivastava}},
  \bibinfo{journal}{Phys. Rev. D} \textbf{\bibinfo{volume}{90}},
  \bibinfo{pages}{045010} (\bibinfo{year}{2014}), \eprint{1405.6492}.

\bibitem[{\citenamefont{Bai and Long}(2018)}]{Bai:2018vik}
\bibinfo{author}{\bibfnamefont{Y.}~\bibnamefont{Bai}} \bibnamefont{and}
  \bibinfo{author}{\bibfnamefont{A.~J.} \bibnamefont{Long}},
  \bibinfo{journal}{JHEP} \textbf{\bibinfo{volume}{06}}, \bibinfo{pages}{072}
  (\bibinfo{year}{2018}), \eprint{1804.10249}.

\bibitem[{\citenamefont{Bai et~al.}(2019)\citenamefont{Bai, Long, and
  Lu}}]{Bai:2018dxf}
\bibinfo{author}{\bibfnamefont{Y.}~\bibnamefont{Bai}},
  \bibinfo{author}{\bibfnamefont{A.~J.} \bibnamefont{Long}}, \bibnamefont{and}
  \bibinfo{author}{\bibfnamefont{S.}~\bibnamefont{Lu}}, \bibinfo{journal}{Phys.
  Rev. D} \textbf{\bibinfo{volume}{99}}, \bibinfo{pages}{055047}
  (\bibinfo{year}{2019}), \eprint{1810.04360}.

\bibitem[{\citenamefont{Asadi et~al.}(2021)\citenamefont{Asadi, Kramer, Kuflik,
  Ridgway, Slatyer, and Smirnov}}]{Asadi:2021yml}
\bibinfo{author}{\bibfnamefont{P.}~\bibnamefont{Asadi}},
  \bibinfo{author}{\bibfnamefont{E.~D.} \bibnamefont{Kramer}},
  \bibinfo{author}{\bibfnamefont{E.}~\bibnamefont{Kuflik}},
  \bibinfo{author}{\bibfnamefont{G.~W.} \bibnamefont{Ridgway}},
  \bibinfo{author}{\bibfnamefont{T.~R.} \bibnamefont{Slatyer}},
  \bibnamefont{and} \bibinfo{author}{\bibfnamefont{J.}~\bibnamefont{Smirnov}}
  (\bibinfo{year}{2021}), \eprint{2103.09822}.

\bibitem[{\citenamefont{Krylov et~al.}(2013)\citenamefont{Krylov, Levin, and
  Rubakov}}]{Krylov:2013qe}
\bibinfo{author}{\bibfnamefont{E.}~\bibnamefont{Krylov}},
  \bibinfo{author}{\bibfnamefont{A.}~\bibnamefont{Levin}}, \bibnamefont{and}
  \bibinfo{author}{\bibfnamefont{V.}~\bibnamefont{Rubakov}},
  \bibinfo{journal}{Phys. Rev.} \textbf{\bibinfo{volume}{D87}},
  \bibinfo{pages}{083528} (\bibinfo{year}{2013}), \eprint{1301.0354}.

\bibitem[{\citenamefont{Huang and Li}(2017)}]{Huang:2017kzu}
\bibinfo{author}{\bibfnamefont{F.~P.} \bibnamefont{Huang}} \bibnamefont{and}
  \bibinfo{author}{\bibfnamefont{C.~S.} \bibnamefont{Li}},
  \bibinfo{journal}{Phys. Rev.} \textbf{\bibinfo{volume}{D96}},
  \bibinfo{pages}{095028} (\bibinfo{year}{2017}), \eprint{1709.09691}.

\bibitem[{\citenamefont{Huber and Konstandin}(2008)}]{Huber:2007vva}
\bibinfo{author}{\bibfnamefont{S.~J.} \bibnamefont{Huber}} \bibnamefont{and}
  \bibinfo{author}{\bibfnamefont{T.}~\bibnamefont{Konstandin}},
  \bibinfo{journal}{JCAP} \textbf{\bibinfo{volume}{05}}, \bibinfo{pages}{017}
  (\bibinfo{year}{2008}), \eprint{0709.2091}.

\bibitem[{\citenamefont{Grojean and Servant}(2007)}]{Grojean:2006bp}
\bibinfo{author}{\bibfnamefont{C.}~\bibnamefont{Grojean}} \bibnamefont{and}
  \bibinfo{author}{\bibfnamefont{G.}~\bibnamefont{Servant}},
  \bibinfo{journal}{Phys. Rev. D} \textbf{\bibinfo{volume}{75}},
  \bibinfo{pages}{043507} (\bibinfo{year}{2007}), \eprint{hep-ph/0607107}.

\bibitem[{\citenamefont{Caprini et~al.}(2016)}]{Caprini:2015zlo}
\bibinfo{author}{\bibfnamefont{C.}~\bibnamefont{Caprini}} \bibnamefont{et~al.},
  \bibinfo{journal}{JCAP} \textbf{\bibinfo{volume}{04}}, \bibinfo{pages}{001}
  (\bibinfo{year}{2016}), \eprint{1512.06239}.

\bibitem[{\citenamefont{Caprini et~al.}(2020)}]{Caprini:2019egz}
\bibinfo{author}{\bibfnamefont{C.}~\bibnamefont{Caprini}} \bibnamefont{et~al.},
  \bibinfo{journal}{JCAP} \textbf{\bibinfo{volume}{03}}, \bibinfo{pages}{024}
  (\bibinfo{year}{2020}), \eprint{1910.13125}.

\bibitem[{\citenamefont{Guo et~al.}(2021)\citenamefont{Guo, Sinha, Vagie, and
  White}}]{Guo:2021qcq}
\bibinfo{author}{\bibfnamefont{H.-K.} \bibnamefont{Guo}},
  \bibinfo{author}{\bibfnamefont{K.}~\bibnamefont{Sinha}},
  \bibinfo{author}{\bibfnamefont{D.}~\bibnamefont{Vagie}}, \bibnamefont{and}
  \bibinfo{author}{\bibfnamefont{G.}~\bibnamefont{White}}
  (\bibinfo{year}{2021}), \eprint{2103.06933}.

\bibitem[{\citenamefont{Jung and Kim}(2020)}]{Jung:2019fcs}
\bibinfo{author}{\bibfnamefont{S.}~\bibnamefont{Jung}} \bibnamefont{and}
  \bibinfo{author}{\bibfnamefont{T.}~\bibnamefont{Kim}},
  \bibinfo{journal}{Phys. Rev. Res.} \textbf{\bibinfo{volume}{2}},
  \bibinfo{pages}{013113} (\bibinfo{year}{2020}), \eprint{1908.00078}.

\bibitem[{\citenamefont{Laha}(2019)}]{Laha:2019ssq}
\bibinfo{author}{\bibfnamefont{R.}~\bibnamefont{Laha}}, \bibinfo{journal}{Phys.
  Rev. Lett.} \textbf{\bibinfo{volume}{123}}, \bibinfo{pages}{251101}
  (\bibinfo{year}{2019}), \eprint{1906.09994}.

\bibitem[{\citenamefont{Dasgupta et~al.}(2020)\citenamefont{Dasgupta, Laha, and
  Ray}}]{Dasgupta:2019cae}
\bibinfo{author}{\bibfnamefont{B.}~\bibnamefont{Dasgupta}},
  \bibinfo{author}{\bibfnamefont{R.}~\bibnamefont{Laha}}, \bibnamefont{and}
  \bibinfo{author}{\bibfnamefont{A.}~\bibnamefont{Ray}},
  \bibinfo{journal}{Phys. Rev. Lett.} \textbf{\bibinfo{volume}{125}},
  \bibinfo{pages}{101101} (\bibinfo{year}{2020}), \eprint{1912.01014}.

\bibitem[{\citenamefont{Laha et~al.}(2020)\citenamefont{Laha, Mu\~noz, and
  Slatyer}}]{Laha:2020ivk}
\bibinfo{author}{\bibfnamefont{R.}~\bibnamefont{Laha}},
  \bibinfo{author}{\bibfnamefont{J.~B.} \bibnamefont{Mu\~noz}},
  \bibnamefont{and} \bibinfo{author}{\bibfnamefont{T.~R.}
  \bibnamefont{Slatyer}}, \bibinfo{journal}{Phys. Rev. D}
  \textbf{\bibinfo{volume}{101}}, \bibinfo{pages}{123514}
  (\bibinfo{year}{2020}), \eprint{2004.00627}.

\bibitem[{\citenamefont{Ray et~al.}(2021)\citenamefont{Ray, Laha, Mu\~noz, and
  Caputo}}]{Ray:2021mxu}
\bibinfo{author}{\bibfnamefont{A.}~\bibnamefont{Ray}},
  \bibinfo{author}{\bibfnamefont{R.}~\bibnamefont{Laha}},
  \bibinfo{author}{\bibfnamefont{J.~B.} \bibnamefont{Mu\~noz}},
  \bibnamefont{and} \bibinfo{author}{\bibfnamefont{R.}~\bibnamefont{Caputo}}
  (\bibinfo{year}{2021}), \eprint{2102.06714}.

\bibitem[{\citenamefont{Lyth and Stewart}(1995)}]{Lyth:1995hj}
\bibinfo{author}{\bibfnamefont{D.~H.} \bibnamefont{Lyth}} \bibnamefont{and}
  \bibinfo{author}{\bibfnamefont{E.~D.} \bibnamefont{Stewart}},
  \bibinfo{journal}{Phys. Rev. Lett.} \textbf{\bibinfo{volume}{75}},
  \bibinfo{pages}{201} (\bibinfo{year}{1995}), \eprint{hep-ph/9502417}.

\bibitem[{\citenamefont{Lyth and Stewart}(1996)}]{Lyth:1995ka}
\bibinfo{author}{\bibfnamefont{D.~H.} \bibnamefont{Lyth}} \bibnamefont{and}
  \bibinfo{author}{\bibfnamefont{E.~D.} \bibnamefont{Stewart}},
  \bibinfo{journal}{Phys. Rev. D} \textbf{\bibinfo{volume}{53}},
  \bibinfo{pages}{1784} (\bibinfo{year}{1996}), \eprint{hep-ph/9510204}.

\bibitem[{\citenamefont{Asaka and Kawasaki}(1999)}]{Asaka:1999xd}
\bibinfo{author}{\bibfnamefont{T.}~\bibnamefont{Asaka}} \bibnamefont{and}
  \bibinfo{author}{\bibfnamefont{M.}~\bibnamefont{Kawasaki}},
  \bibinfo{journal}{Phys. Rev. D} \textbf{\bibinfo{volume}{60}},
  \bibinfo{pages}{123509} (\bibinfo{year}{1999}), \eprint{hep-ph/9905467}.

\bibitem[{\citenamefont{Scherrer and Turner}(1985)}]{Scherrer:1984fd}
\bibinfo{author}{\bibfnamefont{R.~J.} \bibnamefont{Scherrer}} \bibnamefont{and}
  \bibinfo{author}{\bibfnamefont{M.~S.} \bibnamefont{Turner}},
  \bibinfo{journal}{Phys. Rev. D} \textbf{\bibinfo{volume}{31}},
  \bibinfo{pages}{681} (\bibinfo{year}{1985}).

\bibitem[{\citenamefont{Berlin et~al.}(2016{\natexlab{a}})\citenamefont{Berlin,
  Hooper, and Krnjaic}}]{Berlin:2016vnh}
\bibinfo{author}{\bibfnamefont{A.}~\bibnamefont{Berlin}},
  \bibinfo{author}{\bibfnamefont{D.}~\bibnamefont{Hooper}}, \bibnamefont{and}
  \bibinfo{author}{\bibfnamefont{G.}~\bibnamefont{Krnjaic}},
  \bibinfo{journal}{Phys. Lett. B} \textbf{\bibinfo{volume}{760}},
  \bibinfo{pages}{106} (\bibinfo{year}{2016}{\natexlab{a}}),
  \eprint{1602.08490}.

\bibitem[{\citenamefont{Berlin et~al.}(2016{\natexlab{b}})\citenamefont{Berlin,
  Hooper, and Krnjaic}}]{Berlin:2016gtr}
\bibinfo{author}{\bibfnamefont{A.}~\bibnamefont{Berlin}},
  \bibinfo{author}{\bibfnamefont{D.}~\bibnamefont{Hooper}}, \bibnamefont{and}
  \bibinfo{author}{\bibfnamefont{G.}~\bibnamefont{Krnjaic}},
  \bibinfo{journal}{Phys. Rev. D} \textbf{\bibinfo{volume}{94}},
  \bibinfo{pages}{095019} (\bibinfo{year}{2016}{\natexlab{b}}),
  \eprint{1609.02555}.

\bibitem[{\citenamefont{Cosme et~al.}(2021)\citenamefont{Cosme, Dutra, Ma, Wu,
  and Yang}}]{Cosme:2020mck}
\bibinfo{author}{\bibfnamefont{C.}~\bibnamefont{Cosme}},
  \bibinfo{author}{\bibfnamefont{M.~a.} \bibnamefont{Dutra}},
  \bibinfo{author}{\bibfnamefont{T.}~\bibnamefont{Ma}},
  \bibinfo{author}{\bibfnamefont{Y.}~\bibnamefont{Wu}}, \bibnamefont{and}
  \bibinfo{author}{\bibfnamefont{L.}~\bibnamefont{Yang}},
  \bibinfo{journal}{JHEP} \textbf{\bibinfo{volume}{03}}, \bibinfo{pages}{026}
  (\bibinfo{year}{2021}), \eprint{2003.01723}.

\bibitem[{\citenamefont{Kawasaki and Takahashi}(2005)}]{Kawasaki:2004rx}
\bibinfo{author}{\bibfnamefont{M.}~\bibnamefont{Kawasaki}} \bibnamefont{and}
  \bibinfo{author}{\bibfnamefont{F.}~\bibnamefont{Takahashi}},
  \bibinfo{journal}{Phys. Lett. B} \textbf{\bibinfo{volume}{618}},
  \bibinfo{pages}{1} (\bibinfo{year}{2005}), \eprint{hep-ph/0410158}.

\bibitem[{\citenamefont{Leite and Martins}(2011)}]{Leite:2011sc}
\bibinfo{author}{\bibfnamefont{A.~M.~M.} \bibnamefont{Leite}} \bibnamefont{and}
  \bibinfo{author}{\bibfnamefont{C.~J. A.~P.} \bibnamefont{Martins}},
  \bibinfo{journal}{Phys. Rev. D} \textbf{\bibinfo{volume}{84}},
  \bibinfo{pages}{103523} (\bibinfo{year}{2011}), \eprint{1110.3486}.

\bibitem[{\citenamefont{Hattori et~al.}(2015)\citenamefont{Hattori, Kobayashi,
  Omoto, and Seto}}]{Hattori:2015xla}
\bibinfo{author}{\bibfnamefont{H.}~\bibnamefont{Hattori}},
  \bibinfo{author}{\bibfnamefont{T.}~\bibnamefont{Kobayashi}},
  \bibinfo{author}{\bibfnamefont{N.}~\bibnamefont{Omoto}}, \bibnamefont{and}
  \bibinfo{author}{\bibfnamefont{O.}~\bibnamefont{Seto}},
  \bibinfo{journal}{Phys. Rev. D} \textbf{\bibinfo{volume}{92}},
  \bibinfo{pages}{103518} (\bibinfo{year}{2015}), \eprint{1510.03595}.

\end{thebibliography}

\end{document}